\documentclass{ametsocV6.1}

\usepackage[letterpaper,top=2cm,bottom=2cm,left=3cm,right=3cm,marginparwidth=1.75cm]{geometry}

\usepackage{amsmath}
\usepackage{graphicx}
\usepackage[colorlinks=true, allcolors=blue]{hyperref}
\usepackage{soul}
\usepackage{appendix}
\usepackage{caption}
\usepackage{subcaption}
\usepackage{adjustbox}
\usepackage{amssymb}
\usepackage{tikz}
\usepackage{rotating}

\usepackage{float}
\usepackage{array}

\title{  Modeling spatial asymmetries in teleconnected extreme temperatures
}

\authors{Mitchell L. Krock,\aff{1}\correspondingauthor{mkrock@anl.gov}
Julie Bessac,\aff{2,3}
Michael L. Stein \aff{4}
}

\affiliation{\aff{1}{Mathematics and Computer Science Division, Argonne National Laboratory, Lemont, IL, US}\\
\aff{2}{Computational Science Center, National Renewable Energy Laboratory, Golden, CO, US}\\
\aff{3}{Department of Mathematics, Virginia Tech, Blacksburg, VA, US}\\
\aff{4}{Department of Statistics, Rutgers University, New Brunswick, NJ, US}\\
}

\abstract{
    Combining strengths from deep learning and extreme value theory can help describe complex relationships between variables where extreme events have significant impacts (e.g.,\ environmental or financial applications). 
    Neural networks learn complicated nonlinear relationships from large datasets under limited parametric assumptions. 
    By definition, the number of occurrences of extreme events is small, which limits the ability of the data-hungry, nonparametric neural network to describe rare events. 
    Inspired by recent extreme cold winter weather events in North America caused by atmospheric blocking, we examine several probabilistic generative models for the entire multivariate probability distribution of daily boreal winter surface air temperature. 
    We propose metrics to measure spatial asymmetries, such as long-range anticorrelated patterns that commonly appear in temperature fields during blocking events. 
    Compared to vine copulas, the statistical standard for multivariate copula modeling, deep learning methods show improved ability to reproduce complicated asymmetries in the spatial distribution of ERA5 temperature reanalysis, including the spatial extent of in-sample extreme events.}

\begin{document}

\nolinenumbers

\maketitle


\section{Introduction}
\label{sec:introduction}

\begin{figure}[H]
  \centering
  \includegraphics[scale=.24]{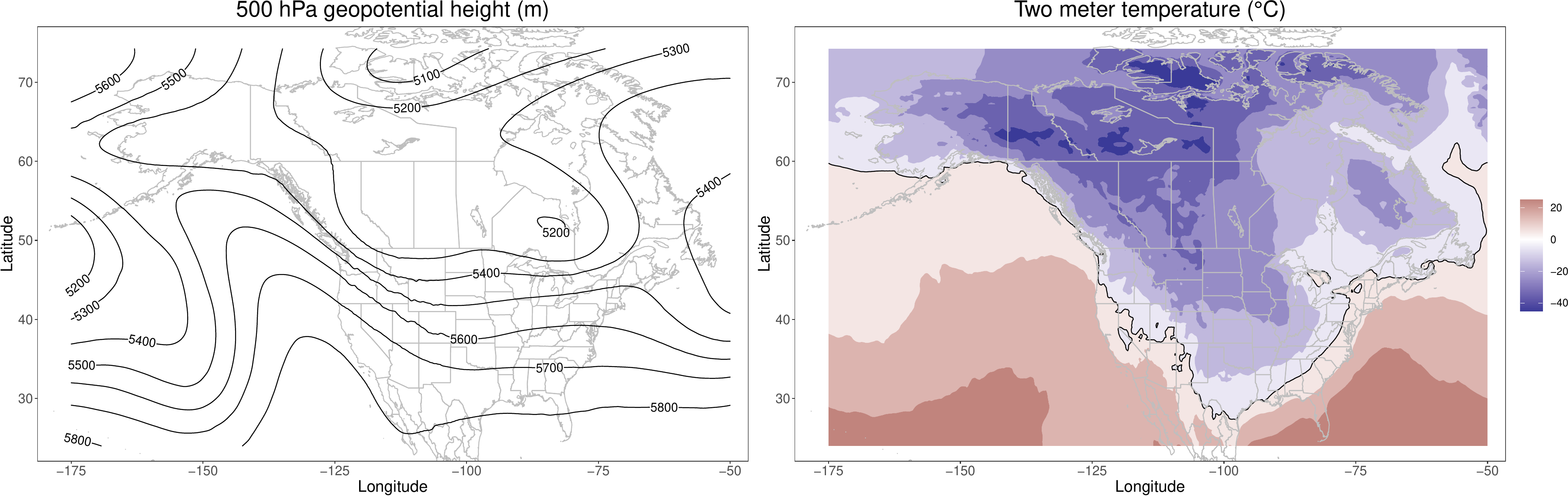}
  \caption{Illustrating the December 2022 winter weather storm with ERA5 data products. The contours in the left figure show the average geopotential height during December 21, 2022.  There is a very large ridge over the NE Pacific and a deep trough over the continental interior.  The flow approaching western coastal Alaska is almost southerly. The right figure shows the average two-meter temperature during December 23, 2022, and the black contour line corresponds to 0$^\circ$C, roughly separating where precipitation falls as rain versus snow.}
  \label{fig:wxmaps}
\end{figure}

In 2022, the U.S.\ experienced a record-setting 18 climate disasters, each causing over \$1 billion in damage \citep{NCEI2022disasters}.
Mitigating risk from environmental extreme events is a pressing issue for science, especially in the face of anthropogenic climate change.
An extreme weather event in late December 2022 motivates the statistical research question investigated in this paper.
During this time, a severe winter storm produced dangerously cold temperatures and significant snowfall in much of the U.S. and Canada.
Buffalo, New York received over 3 feet of snow, and 41 people died during the blizzard.
This storm was the result of an atmospheric blocking, a quasi-stationary high-pressure ridge that disrupts the usual zonal circulation of the atmosphere.
The map of 500 hPa geopotential height in Figure \ref{fig:wxmaps} illustrates that such a blocking event over the subarctic Pacific preluded the 2022 December storm.
Blocking events during the boreal winter in North America are associated with anomalous warmth in Alaska and extreme cold temperatures in the midlatitudes of North America \citep{carrera2004}.
Observe that the temperatures in Figure \ref{fig:wxmaps} are roughly the same in Anchorage, Alaska and El Paso, Texas.

Extant models for spatial extremes are not equipped to deal with teleconnections where one location is abnormally warm while another is abnormally cool \citep{krock2022}.
Specifically, these models are restricted to the joint upper (or lower) tail of the distribution---it is not possible to think about more than one tail of the spatial distribution, which is necessary for opposite-tail teleconnections.
This setting motivates the challenging problem of studying extreme values on large spatial scales while interested in the entire distribution.

We propose to model marginal distributions with the Bulk-And-Tails distribution, a parametric probability distribution with flexible behavior in its upper and lower tails \citep{stein2020}.
After transforming the marginal distributions to standard uniformity, we apply multivariate copula models to capture opposite-tail dependence.
A popular statistical model for this scenario is the vine copula, which uses bivariate building blocks to create a flexible asymmetric multivariate distribution \citep{czado2022}.
However, the simple pairwise construction of vine copulas naturally limits their expressiveness.
We investigate the ability of deep learning models to replicate complicated aspects of the spatial distribution of daily wintertime temperatures in North America. 
In particular, we consider a stochastic process model which combines principal components and normalizing flows.
Compared to vine copulas, the flow-based model more accurately reproduces bivariate opposite-tail teleconnective patterns as well as other spatial patterns in extreme temperatures.

The paper is structured as follows.
In Sections \ref{sec:statbackground} and \ref{sec:mlbackground}, we review concepts from extreme value theory and deep learning, respectively.
Section \ref{sec:related} discusses our model and related work at the intersection of these two research areas.
In Section \ref{sec:era5}, we compare several probabilistic models in a cross-validation study to assess their ability to fit aspects of ERA5 temperature distribution that are relevant to extremes.
Section \ref{sec:discussion} concludes.

\section{Statistical Background}
\label{sec:statbackground}
We review some basic concepts from extreme value analysis, progressing from one variable to two variables to the general multivariate setting.
For a formal introduction, see \citet{coles}.

\subsection{Univariate Extremes}
\label{sec:univariateextremes}
First, we discuss two celebrated univariate models from extreme value theory.
The Fisher-Tippett-Gnedenko theorem motivates using the generalized extreme value (GEV) distribution to model the block-maxima of a random variable, and the Pickands-Balkema-De Haan theorem motivates using the generalized Pareto distribution (GPD) to model threshold exceedances of a random variable.
Both models have explicit parametric control of the tail behavior, which allows researchers to consider ideas such as the ``return level'' of an extreme event (i.e.,\ the expected time until another event that is at least equally extreme).
However, these models only consider a single tail of the distribution, ignoring the bulk and other tail of the data distribution.
\citet{stein2020} proposes a seven-parameter distribution designed for the entire distribution with flexible behavior in both tails.
The cumulative distribution function (cdf) of a Bulk-And-Tails (BATs) random variable $X$ is $P(X \le x) = T_\nu(H(x))$ where $T_\nu(\cdot)$ is the cdf of the student-$t$ distribution with $\nu$ degrees of freedom and $H(\cdot)$ is a monotone increasing function with location, scale, and shape parameters for the upper and lower tails.
The \texttt{BulkAndTails.jl} package provides an interface to this distribution as well as an extension where the location and scale parameters depend on covariates \citep{krock2022bats}.

\subsection{Copulas and Bivariate Tail Dependence}
\label{sec:bivariatetdc}
A copula is a multivariate cdf on the unit hypercube with uniform marginal distributions \citep{Nels06,Joe14}.
For simplicity, consider the bivariate setting where $X$ and $Y$ are two continuous random variables with joint cdf $F_{X,Y}$ and marginal cdfs $F_{X}$ and $F_{Y}$.
Sklar's theorem \citep{sklar1959fonctions} says that there exists a unique copula $C : [0,1] \times [0,1] \to [0,1]$ such that $F_{X,Y}(x,y) = C(F_{X}(x),F_{Y}(y))$.
The copula representation is especially useful in multivariate extreme value statistics, as it provides a straightforward way to construct a valid multivariate distribution while preserving the marginal distributions $F_{X}$ and $F_{Y}$, which presumably have been calibrated to describe the marginal tails of the distribution well.
For the remainder of this paper, we primarily focus on multivariate extremes and assume that the marginal distributions can be modeled appropriately with parametric extreme value models.

\citet{sibuya1960} proposed a tail dependence coefficient to describe the probability of multiple random variables experiencing concurrent extremes.
In the bivariate case, there are four tail dependence statistics \citep{zhang2008}:
\begin{equation} \label{eq:taildependencies}
    \begin{split}
        \lambda_{UU} &=  \lim \limits_{u \uparrow 1} \mathbb{P}(F_{X}(X) > u \mid F_{Y}(Y) > u) = \lim \limits_{u \uparrow 1} \frac{1-2u + C(u,u) }{1 - u}  \\
        \lambda_{LL} &=  \lim \limits_{u \uparrow 1} \mathbb{P}(F_{X}(X) \le 1-u \mid F_{Y}(Y) \le 1-u) = \lim \limits_{u \uparrow  1} \frac{C(1-u,1-u)}{1 - u} \\
        \lambda_{LU} &=  \lim \limits_{u \uparrow 1} \mathbb{P}(F_{X}(X) \le 1-u \mid F_{Y}(Y) > u) = \lim \limits_{u \uparrow 1} \frac{1-u - C(1-u,u) }{1 - u}  \\
        \lambda_{UL} &=  \lim \limits_{u \uparrow 1} \mathbb{P}(F_{X}(X) > u \mid F_{Y}(Y) \le 1-u) = \lim \limits_{u \uparrow 1} \frac{1-u - C(u,1-u) }{1 - u}.
    \end{split}
\end{equation}

Applications usually focus on $\lambda_{UU}$ and $\lambda_{LL}$, yet $\lambda_{UL}$ and $\lambda_{LU}$ have received attention in financial time series \citep{WANG20131706,chang2021}.
\citet{krock2022} studied spatial patterns of these four tail dependencies using ERA5 reanalysis of winter surface air temperature.
Atmospheric blocking over the subarctic Pacific produces strong teleconnections between Alaska and the midlatitudes of North America, which is reflected in large values for opposite-tail dependencies $\lambda_{UL}$ and $\lambda_{LU}$ (illustrated later in Figure \ref{fig:maptaildependencies}). 
Teleconnections are traditionally studied with correlation coefficients \citep{wallace1981}. 
Tail dependence coefficients provide an alternative way to measure the strength of teleconnections when one is interested in extremes.

With these details in mind, the main question is what copula to use.
Two common families of copula models are elliptical copulas, which are highly symmetric, or Archimedean copulas, which only possess a single parameter, restricting their utility to low dimensions.
These two families also have limitations on the possible values for tail dependencies; in particular, none are capable of producing four distinct values in \eqref{eq:taildependencies}.
The bivariate mixture model proposed in \citet{krock2022} considers a mixture of four $90^\circ$ rotations of a bivariate Archimedean copula with asymptotic dependence in one corner of the unit square and asymptotic independence in the other three corners. 
By construction, this model has four different values for the tail dependence coefficients \eqref{eq:taildependencies}, but their sum cannot exceed 1.
This limitation is unrealistic since $\lambda_{UU}$ and $\lambda_{LL}$ will both be close to 1 when considering two nearby locations.
Moreover, extending this mixture model to $d > 2$ dimensions is unrealistic, as it requires estimating a weight and a copula parameter for each $d$-dimensional tail, yielding a total of $2^d$ parameters.

We may also be interested in the values in \eqref{eq:taildependencies} at a quantile $u < 1$, in which case nonparametric empirical estimators are available.
For example, given independent and identically distributed realizations $(x_1,y_1),\dots,(x_n,y_n)$ of $(X,Y)$, \citet{reiss2007} define
\begin{equation}
  \label{eq:empiricaltaildependence}
  \hat \lambda_{UU}(u) = \frac{1}{n(1-u)} \sum_{i=1}^n \mathbf{1} \left( x_i > x_{\lfloor un \rfloor : n} \text{ and } y_i > y_{\lfloor un \rfloor : n} \right)
\end{equation}
where $x_{\lfloor un \rfloor : n}$ is the $\lfloor un \rfloor$ largest value among $(x_1,\dots,x_n)$ and $\mathbf{1}(\cdot)$ is an indicator function that equals one when its argument is true and zero otherwise. 
Note that, for a fixed amount of data, the quality of these empirical estimates will worsen as $u$ is increased to 1.

\subsection{Multivariate Extremes}
\label{sec:multivariatetdc}

\citet{li2009} generalizes the bivariate tail dependencies \eqref{eq:taildependencies} to arbitrary orthants of a $d$-dimensional vector.
When working with spatial data from a large study region, it can be unrealistic to only consider scenarios where all $d$ variables are simultaneously extreme.
In particular, for long-range anticorrelated temperature extremes, there must be a transition region of asymptotic independence between the two regions that experience opposite-tail extremes.

Flexible multivariate copulas with analytic expressions for tail dependencies are scarce.
In this work, we take a unique approach and validate our models through comparison with empirical tails of the data distribution.
Assessing model fit based on empirical tail dependence coefficients is already challenging due to an inherent lack of data, and this effect is compounded when one is interested in multivariate opposite-tail extremes.
Later in Section \ref{sec:era5}, we propose metrics based on the ``spatial extent'' of an extreme event to assess the strength of opposite-tail dependence.

Besides vine copulas (which we discuss next), there are few other options for constructing asymmetric multivariate copulas.
Archimax copulas combine Archimedean copulas with a stable tail dependence function for added flexibility \citep{charpentier2014}.
\citet{ng2022} developed scalable methods for inference and sampling from the Archimax family and demonstrate improved performance in tail dependence inference compared to several other density estimators.
\citet{gong2022asymmetric} proposed a copula model with asymmetric behavior in the joint upper and lower tails, but computations in high dimensions are demanding.
A main limitation of multivariate copula models outside the vine family lies in their restriction to capturing tail dependence and asymmetries along the ``main diagonal'' of the distribution.
That is, they are only concerned with modeling extremes in the joint upper (lower) tail, which effectively ignores any sort of negative extremal dependence in the distribution. 

Existing models in spatial extremes suffer from the same fundamental limitation.
Max-stable processes and generalized Pareto processes generalize the univariate GEV and GPD methodology to the setting of multivariate stochastic processes.
Both models are restricted to the joint upper (or lower) tail of the distribution, and the associated tail dependence coefficient is constant over space.
Many recent works have proposed nonstationary models that can transition from close-range asymptotic dependence to asymptotic independence as the distance between locations increases; see \citet{huser2022} for a review.
Teleconnections can cause tail dependence to appear at large distances, potentially between opposite tails of the distribution. 
Conditional extremes models \citep{heffernan2004} could be adapted to deal with teleconnections, but this would be a significant departure from the current framework where the conditioning extreme event only considers the joint upper tail, and often just a single conditional site \citep{wadsworth2022}.

\subsection{Vine Copulas} 
\label{sec:vinecopula}
In general, it is not possible to construct a multivariate distribution that preserves a collection of bivariate marginal distributions \citep[Section 2.7]{Joe14}.
We can, however, approximate this procedure with vine copulas \citep{bedford2002, aas2009, czado2022}, which construct a valid multivariate copula using $d(d-1)/2$ bivariate copulas as building blocks.
Specifically, the $d$-dimensional copula is factorized as a product of $d(d-1)/2$ bivariate copulas according to a vine-like conditional independence structure.
Commonly-used canonical vines (C-vines) or drawable vines (D-vines) are special cases of the regular vine (R-vine) family.
Given a regular vine $V$ with edge set $E$, an R-vine copula density expresses as
\begin{align}
  \label{eq:vinecopula}
  f(x_1,x_2,\dots,x_d) = \prod_{e \in E} c_{e_1,e_2;D_e} (F_{e_1}|D_e , F_{e_2}|D_e ; D_e ) \prod_{i=1}^d f_{i}(x_i)
\end{align}
where $D_e$ denotes the conditioning variables corresponding to the edge $e = (e_1, e_2)^{\mathrm{T}}$.
An example of a four-dimensional R-vine copula density given by \citet{czado2022} is
\begin{align}
  \label{eq:vinecopula4}
  f(x_1,x_2,x_3,x_4) &= c_{12}(F_1(x_1), F_2(x_2)) \times c_{13}(F_1(x_1), F_3(x_3)) \times c_{14}(F_1(x_1), F_4(x_4)) \nonumber \\
  & \times c_{23;1}(F_{2|1}(x_2|x_1), F_{3|1}(x_3|x_1); x_1) \times c_{24;1}(F_{2|1}(x_2|x_1), F_{4|3}(x_4|x_1); x_1) \\
  & \times c_{34;12}(F_{3|12}(x_3|x_1, x_2), F_{4|12}(x_4|x_1, x_2); x_1, x_2) \times \prod_{i=1}^4 f_{i}(x_i).  \nonumber
\end{align}
We see that the joint density of a vine copula framework factorizes into three parts: the marginal distributions, a set of baseline copulas that connect the marginal distributions (i.e.,\ $c_{12}, c_{13}, c_{14}$), and conditional copulas that use the previous edges as leaves (i.e.,\ $c_{23;1}, c_{24;1}, c_{34;12}$).
\citet{joe2010} showed that these baseline copulas govern the tail dependence of the vine copula.
In total, there are $d! \times 2^{ (d-2)(d-3)/2-1}$ R-vine copulas in $d$ dimensions \citep{morales}.

Vine copulas are widely used in finance \citep{brechmann2013,low2018} and have also appeared in spatial statistics literature \citep{graler2014,erhardt2015}.
Tail dependence coefficients of a vine copula can be calculated recursively \citep{joe2010, SalazarFlores2021}.
In particular, \citet{SalazarFlores2021} use rotated copulas to extend the recursive derivations of tail dependence functions in \citet{joe2010} to the setting of counterdiagonal/nonpositive dependence; i.e.,\ they consider an arbitrary joint tail where all $d$ variables are marginally extreme, but not necessarily the joint upper or lower tail.
Except in special cases, this recursion will require numerical computation of high-dimensional integrals.
Instead of approximating this integral with repeated Monte Carlo integration, we take a simpler approach and estimate the tail dependence coefficients empirically from simulations from the R-vine copula \citep[Algorithm 2.2]{dissmann2013}.
\citet{joe2010} also showed that a vine copula has tail dependence in the joint upper tail if each $d-1$ bivariate copula in the first baseline level of the vine exhibits upper tail dependence.
This reasoning extends directly to other counterdiagonal tails by rotating copulas as in \citet{SalazarFlores2021}.
That is, tail dependence exists in one of the $2^d$ joint tails if the $d-1$ bivariate baseline copulas are extreme in appropriate corners of the unit square.
For example, if we desire nonzero dependence in the $\lambda_{ULLU}$ tail of \eqref{eq:vinecopula4}, then the bivariate copula densities $c_{12}, c_{13}$, and $c_{14}$ must have nonzero tail dependence in the $\lambda_{UL}$, $\lambda_{UL}$ and $\lambda_{UU}$ corners of the unit square, respectively. 
Thus, considering arbitrary $d$-dimensional tail dependencies in vine copula models can be difficult since the typical bivariate building blocks do not have flexible tail dependencies in the four corners of the unit square.
The relationship between the $d$-dimensional joint tails and lower-dimensional tails of a vine copula is complicated; see Proposition 4.3 in \citet{joe2010} and \citet{simpson2021}. 

In summary, modeling opposite-tail teleconnected extremes  motivates us to search for multivariate distributions with flexible behavior in multiple tails---not just the joint upper tail, which is the setting of nearly all research in multivariate extreme value theory.
Vine copulas are suited for this task, but ensuring nonzero tail dependence in a situation where marginal distributions are mixed between the upper tail, lower tail, and bulk is outside the scope of current literature.


\section{Deep Learning Background}
\label{sec:mlbackground}
We explore the ability of probabilistic generative models to model complex features of the entire data distribution, with particular focus on asymmetries and extremes.
A probabilistic generative model in this context refers to a model which is able to generate unconditional samples from a multivariate probability distribution.

\subsection{Normalizing Flow}
\label{sec:normalizingflows}
A normalizing flow uses neural networks to parameterize an invertible map from a simple known distribution to a complex unknown data distribution.
Therefore, normalizing flows can be applied in Bayesian settings as prior distributions \citep{Rezende2015}.
Specifically, a normalizing flow $g_\theta: \mathbb{R}^\ell \to \mathbb{R}^\ell$ is an invertible function with an easily-computable Jacobian determinant such that the transformed random variable $\mathbf{Z} = g_\theta^{-1}(\mathbf{C})$ follows a simple base distribution (e.g.,\ $\mathbf{Z} \sim N(\mathbf{0},I_\ell)$ where $I_\ell$ is the $\ell \times \ell$ identity matrix).
Using the change-of-variables formula, the density
\begin{equation}
  \label{eq:nfdensity}
  f_\mathbf{C}(\mathbf{c}) = f_\mathbf{Z}(g_\theta^{-1}(\mathbf{c})) \left| \text{det} \left( \frac{d g_\theta(\mathbf{c})}{d \mathbf{c}} \right) \right|^{-1}
\end{equation}
is trivial by construction.
Note that a composition of normalizing flows is still invertible with tractable Jacobian determinant, so in practice, it is common to compose normalizing flows for increased modeling flexibility.
We follow standard training procedure of neural network models and repeatedly update estimates for $\theta$ using stochastic gradient descent where the objective function (likelihood of the model) is evaluated using a subset of the data samples (known as a ``batch'').
Training happens for a number of epochs, where an epoch is a pass through all batches, updating $\theta$ via stochastic gradient descent in each batch. 
To prevent overfitting, batches are shuffled randomly between epochs.
In our implementation, we use the state-of-the-art Neural Spline Flow \citep{Durkan2019} implemented in the Python package \texttt{nflows} \citep{nflows}.
The Neural Spline Flow is a powerful density estimator, particularly when constructed in an autoregressive fashion \citep{coccaro2023curse}.
We provide a more detailed description of the autoregressive Neural Spline Flow in Appendix \ref{app:normalizingflows}; see \citet{Kobyzev2021} and \citet{Papamakarios2022} for extensive discussion of normalizing flows.
A downside of this model is that the user is required to select an autoregressive ordering.
In Section \ref{sec:related}, we propose a solution based on principal components to help with this problem of ordering variables.

\subsection{Other Generative Models}
\label{sec:othergenerativemodels}
To produce a new simulation from the data distribution, one follows the flow from the noise distribution to the data distribution (i.e.,\ $\mathbf{Z} \mapsto g_\theta(\mathbf{Z})$).
However, this does not require invertibility of the generator $g_\theta(\cdot)$.
Instead of a normalizing flow, one could use a standard feedforward neural network $g_\theta : \mathbb{R}^L \to \mathbb{R}^\ell$ with $L \ge \ell$ for increased modeling flexibility.
To be precise, this $g_\theta(\cdot)$ would be defined recursively via $J$ hidden layers as 
\begin{equation}
  \label{eq:mlp}
  g_\theta(\cdot) := g^{(J)}(\cdot) = \sigma(W_J g^{(J-1)}(\cdot) + \mathbf{b}_J)
\end{equation}
where $\sigma(\cdot)$ is a nonlinear activation function applied elementwise to its argument, $g^{(0)}(\cdot)$ is the identity function, and the parameters $\theta$ are the weight matrices $W_1,\dots,W_J$ and bias vectors $\mathbf{b}_1,\dots,\mathbf{b}_J$.
For simplicity, we set $L = \ell$ in this paper and demonstrate that the invertible normalizing flow framework performs as well as the noninvertible framework \eqref{eq:mlp}.

If $g_\theta(\cdot)$ is not invertible, the likelihood \eqref{eq:nfdensity} is unknown and cannot be used as the objective function of the model.
Some alternatives to maximum likelihood are adversarial training \citep{gan} or discrepancy training \citep{gmmn}.
Although generative adversarial networks (GANs) have achieved widespread success in many applications, adversarial training is known to be unstable, and we encountered training difficulties in our data analysis (see Appendix \ref{app:synthetic}). 
\citet{annau2023algorithmic} show that GANs can hallucinate small-scale artifacts within downscaled model predictions of surface wind.
We found discrepancy metrics based on proper scoring rules to be a more effective than adversarial training.
Note that maximum likelihood estimation is equivalent to minimizing a distance---the KL-divergence---between the data distribution and model distribution. 
The \texttt{GeomLoss} package \citep{feydy2019interpolating} provides a PyTorch loss function that calculates the energy distance between distributions.
Formally, the squared energy-distance between two distributions $F_A$ and $F_B$ equals $
2 \mathbb{E} \| \mathbf{A} - \mathbf{B} \| - \mathbb{E} \|\mathbf{B} - \mathbf{B}'\| - \mathbb{E} \| \mathbf{A}- \mathbf{A}'\|$, where $\mathbf{A}_1,\mathbf{A}_2, \mathbf{B}_1,\mathbf{B}_2$ are independent with $\mathbf{A}_1,\mathbf{A}_2 \sim F_A$ and $\mathbf{B}_1,\mathbf{B}_2 \sim F_B$.
To minimize this distance in practice, the expectations are approximated with the sample mean, and neural network parameters in the generator are optimized with stochastic gradient descent.
These alternative training procedures are also available for normalizing flows \citep{si2021,si2022}, but for the remainder of this paper, our normalizing flow models are estimated by maximum likelihood. 

\section{Proposed Model and Related Work}
\label{sec:related}

To model dependence between variables, we consider the process model $Y(\mathbf{s}) = \sum_{i=1}^\ell C_i \phi_i(\mathbf{s})$ where the random vector $\mathbf{C} = (C_1,\dots,C_\ell)^{\mathrm{T}}$ is modeled using the Neural Spline Flow \citep{Durkan2019} under different choices of basis functions $\phi_1,\dots,\phi_\ell$ with $\ell \le d$.
Given temperature measurements at locations $\mathbf{s}_1,\dots,\mathbf{s}_d$, we are interested in the random vector $\mathbf{Y}$ where
\begin{equation}
  \label{eq:basisflow}
  \mathbf{Y} = 
  \begin{pmatrix}
    Y(\mathbf{s}_1) \\
    Y(\mathbf{s}_2) \\
    \vdots \\
    Y(\mathbf{s}_d) \\
  \end{pmatrix}
  = \begin{pmatrix}
    \phi_1(\mathbf{s}_1) & \phi_2(\mathbf{s}_1) & \dots & \phi_\ell(\mathbf{s}_1) \\
    \phi_1(\mathbf{s}_2) & \phi_2(\mathbf{s}_2) & \dots & \phi_\ell(\mathbf{s}_2) \\
    \vdots & & \ddots & \vdots \\
    \phi_1(\mathbf{s}_d) & \phi_2(\mathbf{s}_d) & \dots & \phi_\ell(\mathbf{s}_d) \\
  \end{pmatrix}
  \begin{pmatrix}
    C_1 \\
    C_2 \\
    \vdots \\
    C_\ell
  \end{pmatrix} 
  = \Phi \mathbf{C}.
\end{equation}
Here, $\mathbf{Y}$ is implicitly assumed to have standard normal marginals, although this is not enforced in our model.
We build basis functions from principal component analysis (PCA), a popular statistical technique for dimension reduction.
Constructing $\Phi$ from principal components has been effective in other normalizing flow applications \citep{cunningham2020normalizing, cramer2022,ijcai2022p448,klein2022}, although none of these works consider spatial processes or extremes.
\citet{jiang2020} and \citet{drees2021} conduct principal component analyses for multivariate extremes with respect to the joint upper tail of the distribution.
We can also model $\mathbf{Y}$ directly with the normalizing flow; i.e.,\ taking $\ell = d$ and $\Phi= I_\ell$.
In general, basis function approaches for spatial data can be useful when number of locations is prohibitively large.
Using principal components also adds a notion of spatial continuity which is otherwise missing in the normalizing flow model.
An additional motivation in our data analysis is that the first principal component shows a pronounced opposite-tail teleconnection between Alaska and the rest of the study region (see Figure \ref{fig:eof}).

Assuming the number of principal components is large enough, the density of \eqref{eq:basisflow} can be approximated as
\begin{equation}
  \label{eq:basisflowdensity}
  f_\mathbf{Y}(\mathbf{y}) \approx f_\mathbf{C}(\Phi^\dagger \mathbf{y}) \left \vert \text{det}\left(\Phi^\dagger \Phi \right) \right \vert^{-1/2}
\end{equation}
where $\Phi^\dagger$ is the left-pseudoinverse of $\Phi$ and $f_\mathbf{C}(\cdot)$ is the normalizing flow density.
Using a normalizing flow instead of an arbitrary probabilistic generative model ensures that $f_\mathbf{C}(\cdot)$ is available in closed form.
Observe that the neural network parameters in \eqref{eq:basisflowdensity} only depend on the first term $f_\mathbf{C}(\Phi^\dagger \mathbf{y})$, and the projected data $\Phi^\dagger \mathbf{y}$ can be precomputed.

As a model for the entire distribution of multivariate extremes, we use the Bulk-And-Tails distribution for marginal distributions in conjunction with a normalizing flow for the dependence structure.
This framework is a special case of a Copula and Marginal (CM)-Flow \citep{wiese2019}, allowing for flexible multivariate tail modeling up to an approximation of the uniform distribution.
It is important to note that an arbitrary normalizing flow used in the copula step of a CM-flow is not a true copula model, as its marginal distributions are unknown.
However, once training is complete, the marginal distributions can be corrected to more closely resemble a copula in a post-processing step.
Our modeling procedure is as follows:
\begin{enumerate}
  \item \textbf{Marginals}: At each of the $d$ locations, fit the observations with a BATs distribution. We allow the scale and location parameters to depend on time-varying covariates to account for differences between temperatures at the beginning and end of boreal winter \citep{krock2022bats}.
  \item \textbf{Copula}: Transform marginal distributions to standard normality by applying the estimated BATs cdf followed by the normal quantile function. If desired, project the data onto its principal components; the number of principal components can be chosen according to the percentage of variability explained. The generative model is then fit to the (projected) data by maximizing the likelihood \eqref{eq:basisflowdensity}.
  \item \textbf{Correction}: Simulate from $n_{\text{gen}}$ times from the $d$-dimensional model, with $n_{\text{gen}} > n_{\text{dat}}$. For each marginal distribution, estimate a univariate cdf based on the simulations (e.g.,\ empirical cdf or parametric BATs cdf). Then, for each marginal distribution, apply the estimated cdf followed by the normal quantile function.
\end{enumerate} 
A copula can equivalently be defined with marginal distributions that are standard normal instead of standard uniform. 
It is more natural to use the convention of standard normal marginals than to constrain the basis expansion \eqref{eq:basisflow} to lie in the unit hypercube.
For Step 3, we can consider parametric and nonparametric corrections for the marginals of the generative model.
Ideally, the estimated marginal distributions of the generative model are already close to standard normal, so only minor corrections are needed.
Note that applying a strictly-increasing parametric function (e.g.,\ the cdf of the BATs distribution) to correct marginals is guaranteed to preserve the copula \citep{Joe14}. 

\subsection{Related work}

\cite{boulaguiem2022} use a similar framework to model the annual maxima of temperature and precipitation in western Europe.
Specifically, their model combines GEV marginals with a GAN for dependence.
The key difference is that this formulation only considers the joint upper tail of the data.
If the GEV marginals in \cite{boulaguiem2022} were changed to the BATs distribution, it would look similar to our model but with the normalizing flow replaced by a GAN.
We found this GAN version of our model very difficult to train on the ERA5 dataset (see Section \ref{app:synthetic}).
However, with gridded climate data, \cite{boulaguiem2022} were able to use convolutional layers in the GAN architecture, which could explain their improved performance.

The closest work to ours is \citet{mcdonald2022}, who raise the important issue of creating a generative multivariate model with flexible tail dependence.
They propose COMET Flows, a specific type of CM-flow which combines a normalizing flow for dependence with flexible marginal models for extremes.
Their marginal distributions are obtained from a mixture model that combines a kernel density estimate for the bulk of the distribution with two generalized Pareto distributions to describe the lower and upper marginal tails.
This type of model has been explored in \citet{scarrott2011}; see \citet{scarrott2012} for extensive discussion about threshold models and issues with combining bulk and tails in inference for extremes.

With the marginals specified, \citet{mcdonald2022} use SoftFlow \citep{Kim2020} to model the dependence between variables.
The motivation for using SoftFlow is that tail dependence can be viewed as a low-dimensional manifold of feature space that normalizing flows struggle to model since they are invertible.
SoftFlow perturbs conditional inputs with noise to better capture the manifold structure in feature space.
Despite this justification, we did not find SoftFlow to be an effective solution to modeling multivariate tail dependence (see Section \ref{app:synthetic}).
In comparison to \citet{mcdonald2022}, we focus explicitly on spatial patterns of extremes and consider the basis decomposition \eqref{eq:basisflow} to move towards a stochastic process model.

\subsection{Marginal Tails}
\label{sec:marginals}
Although this paper focuses on multivariate tails of the distribution, we briefly mention some important results about the marginal tails of probabilistic generative models.
Previous work has studied how the marginal tails of the data distribution and the base distribution are related, both for normalizing flows \citep{jaini2020,wiese2019,laszkiewicz2022} and GANs \citep{huster2021, oriol2021, allouche2022}.
If, for example, marginals of the data distribution are heavy-tailed but the base distribution is Gaussian noise, then common GANs or normalizing flows will fail. 
The flow used in our experiments does not possess the limitations described in these works; see Appendix \ref{app:normalizingflows} for more discussion. 
Moreover, we marginally transform the copula to match the standard Gaussian marginals of the base distribution so that, in theory, the marginal modeling step is less difficult.

\section{ERA5 Data Analysis}
\label{sec:era5}
To illustrate the strengths of our proposed multivariate model, we build on the data analysis from \citet{krock2022} that used tail dependence coefficients to study teleconnected extremes in ERA5 temperature reanalysis data product \citep{ERA5}. 
Tail dependence coefficients provide an alternative way to measure the strength of teleconnected extremes beyond the commonly-used correlation coefficient \citep{wallace1981}.
As discussed in Section \ref{sec:statbackground}\ref{sec:multivariatetdc}, generalizing the bivariate mixture model from previous work to a high-dimensional setting is unrealistic.
We propose several criterion for model performance and conduct a cross-validation study among various probablistic generative models to see how well they model aspects of the temperature distribution that are relevant to extremes.

We consider daily average two-meter temperature in December, January, and February from 1979-2022 over Canada and the contiguous United States; in total, there are $n_\text{dat}=3940$ temperature measurements at each location.
The original ERA5 product lies on a 0.25 degree longitude/latitude grid, and we perform a coarse spatial averaging, resulting in $d=76$ disjoint gridboxes that cover the study region.
For the remainder of this work, we consider marginal modeling at each of the $76$ locations to be complete after fitting marginal BATs models \citep{krock2022bats} and transforming the temperature data to have standard uniform (or normal) marginal distributions.
That is, we assume temporal independence over days after applying the marginal transformations to uniformity, giving us samples $\mathbf{Y}_1,\dots,\mathbf{Y}_{n_\text{dat}}$ of the $d$-dimensional random vector $\mathbf{Y}$ with known marginals.
Although incorporating time dependence is necessary for predicting when an atmospheric blocking will occur, we focus on the spatial distribution of the temperature field during blocking events, for which the notion of simultaneous extremes is more important.

\begin{figure}[H]
  \centering
  \includegraphics[scale=.4]{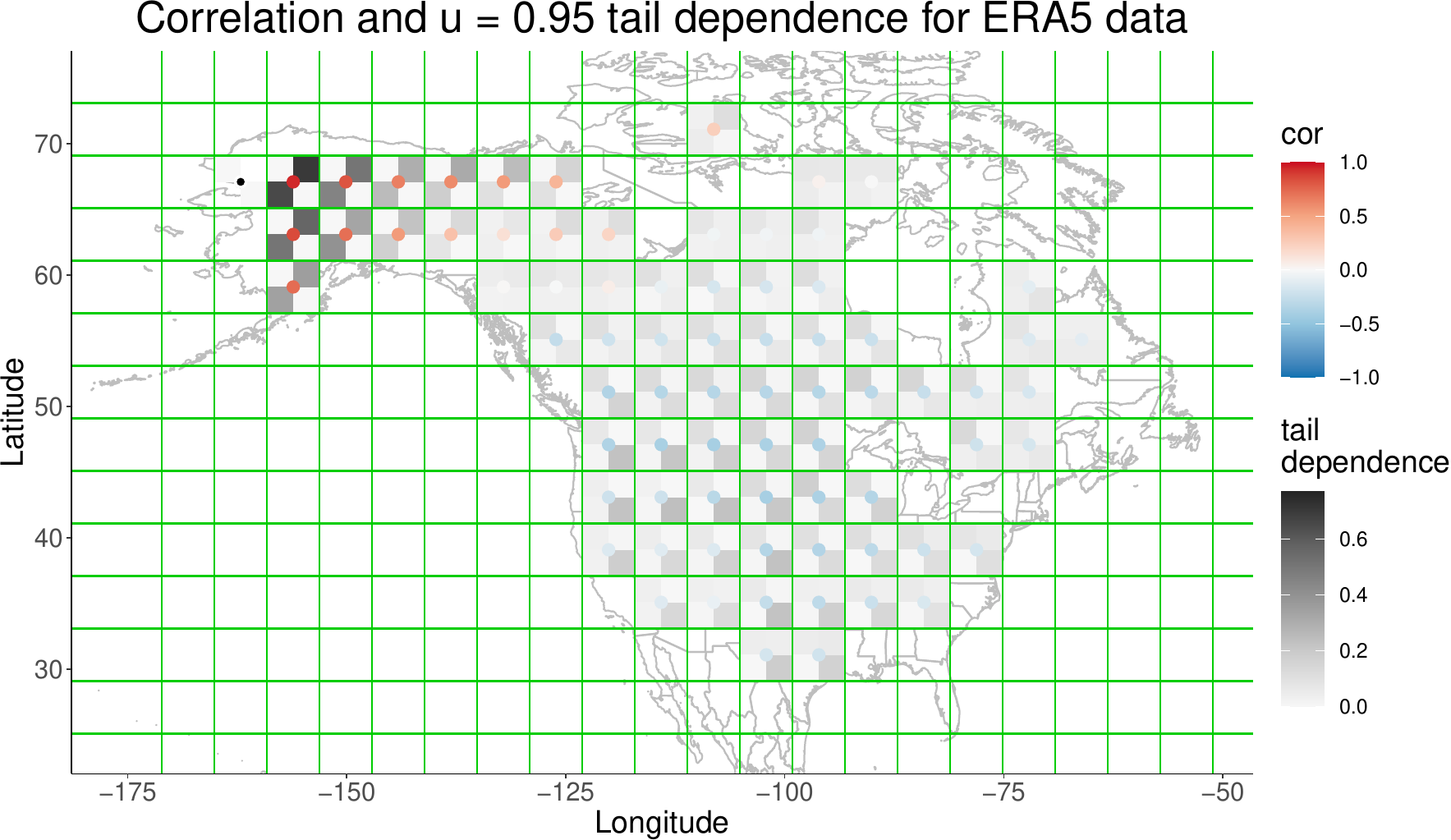}
  \caption{
  Correlation and empirical $u=0.95$ tail dependencies between gridbox in NW Alaska (black dot) and all other pairs of gridboxes. For reference, the black dot  corresponds to a gridbox in NW Alaska whose lower left coordinate is $(-165.1^\circ, 65.1^\circ)$ and upper right coordinate is $(-159.1^\circ, 69.1^\circ)$.
    Grayscale denotes the gridbox values of the extremal dependence matrix (such that the lower right corner of a gridbox depicts $\lambda_{UL}$, the tail dependence where that gridbox is especially cold and the gridbox with the black dot is especially warm.)}
  \label{fig:maptaildependencies}
\end{figure}

Figure \ref{fig:maptaildependencies} shows empirical estimates of the four tail dependencies \eqref{eq:taildependencies} and bivariate correlation between a gridbox in NW Alaska and all other 75 gridboxes.
Spatial patterns are evident, especially the strong opposite-tail dependence between NW Alaska and mid-latitude U.S. corresponding to atmospheric blocking events.
Near the primary gridbox in NW Alaska, there is strong positive dependence in both the correlation coefficient and the common-tail dependencies $\lambda_{LL}$ and $\lambda_{UU}$.
Moving towards northern Canada, both the correlation and tail dependence coefficients are small.
Over most of Canada and the U.S.,\ we see teleconnective patterns of negative correlation and large values for opposite-tail dependence.
The strongest teleconnections correspond to atmospheric blocking events where the U.S.\ experiences freezing temperatures while NW Alaska is anomalously warm.
In contrast, for central Canada, the larger opposite-tail dependence coefficient corresponds to zonal atmospheric flow where Canada is relatively warm while Alaska is relatively cool.
These asymmetries also exist in the analytic tail dependencies for parametric bivariate mixture models in \citet{krock2022}.
A main goal of our multivariate model in this paper is to reproduce these bivariate correlations and (empirical) tail dependencies.
Spatial processes based on elliptical distributions can model a wide range of correlations but have opposite-tail symmetry in their bivariate tail dependence coefficients.

As a way to study opposite-tail dependence beyond the bivariate case, we propose statistics based on the spatial extent of an extreme event, defined in \citet{Zhang2022}.
First, in a preliminary step, we calculate the areas of the $d=76$ coarse gridboxes using great circle distance\footnote{Gridbox areas are calculated using great circle distance: $(\pi /180) R^2  | \sin(\text{lat}_1) - \sin(\text{lat}_2) |  | \text{lon}_1 - \text{lon}_2 |$ where $R = 6371$ km is the radius of the Earth and $[\text{lon}_1, \text{lat}_1] \times [\text{lon}_2, \text{lat}_2]$ is the boundary of the gridbox.} and call them $A_1,\dots,A_d$.
For a temperature vector $\mathbf{U} = (U_1,\dots,U_d)^{\mathrm{T}}$ on the copula scale, we define the statistics  
\begin{equation} \label{eq:spatialextent}
  \begin{split}
      \alpha_{UU} &=  \lim \limits_{u \uparrow 1} \mathbb{E} \left( \sqrt{\sum_{i=1}^d A_i \mathbf{1}(U_i > u) / \pi } \mid U_j > u \right)\\
      \alpha_{LL} &=  \lim \limits_{u \uparrow 1} \mathbb{E} \left( \sqrt{\sum_{i=1}^d A_i \mathbf{1}(U_i \le 1-u) / \pi } \mid U_j \le 1-u \right)\\
      \alpha_{LU} &= \lim \limits_{u \uparrow 1} \mathbb{E} \left( \sqrt{\sum_{i=1}^d A_i \mathbf{1}(U_i \le 1-u) / \pi } \mid U_j > u \right)\\
      \alpha_{UL} &=  \lim \limits_{u \uparrow 1} \mathbb{E} \left( \sqrt{\sum_{i=1}^d A_i \mathbf{1}(U_i > u) / \pi } \mid U_j \le 1-u \right)
  \end{split}
\end{equation}
to represent the average radius\footnote{Note that $\sqrt{A/\pi}$ is the radius of a circle with area $A$.} of exceedance (ARE) of spatial extremes in $\mathbf{U}$ given that $U_j$ is extreme.
Our extension to \citet{Zhang2022} is to consider cases other than the upper-tail extent $\alpha_{UU}$.

We empirically estimate the tail dependencies \eqref{eq:taildependencies} and ARE \eqref{eq:spatialextent} with samples from the probabilistic generative model.
Since our primary interest is accurately modeling observed extreme events rather than extrapolation, we use a large number of simulations (i.e.,\ $n_{\text{gen}} = 10^6$) and select a modest quantile level of $u=0.95$ to define empirical extremes.
Increasing $u$ further suggests that none of estimated generative models exhibit long-range opposite-tail dependence (see Figure \ref{app:farthertails}).
By setting aside analytic expressions for \eqref{eq:taildependencies} and \eqref{eq:spatialextent}, we ignore the issue of extrapolating beyond the in-sample behavior, which is a primary role of extreme value theory.
Nonetheless, the spatial asymmetries of moderately extreme surface air temperatures in our data example pose an interesting and relevant challenge outside the usual scope of spatial extremes methodology.



\subsection{Model Comparison}
\label{sec:modelcomparison}

\begin{figure}[H]
  \centering
  \includegraphics[scale=.23]{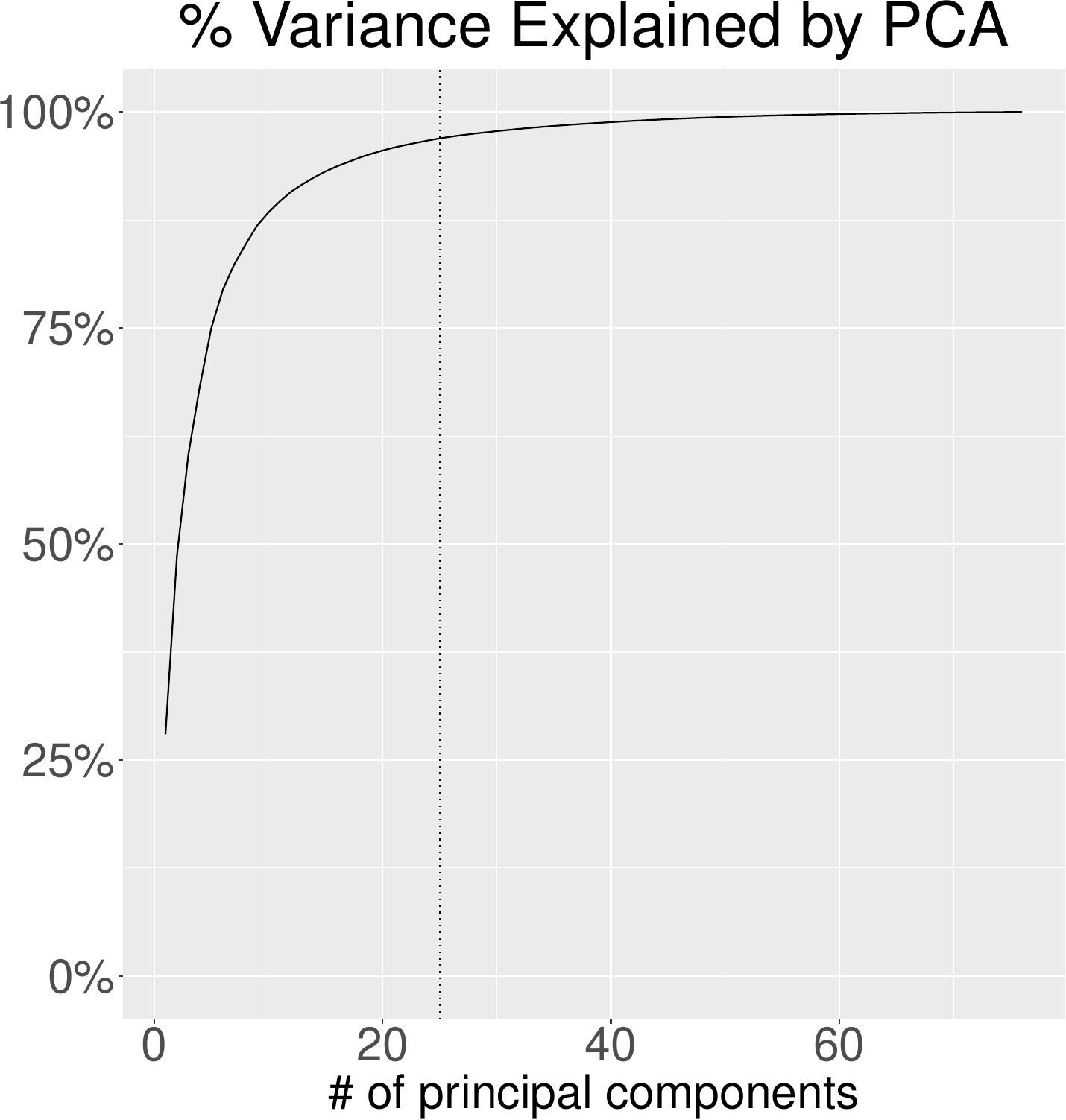}
  \includegraphics[scale=.33]{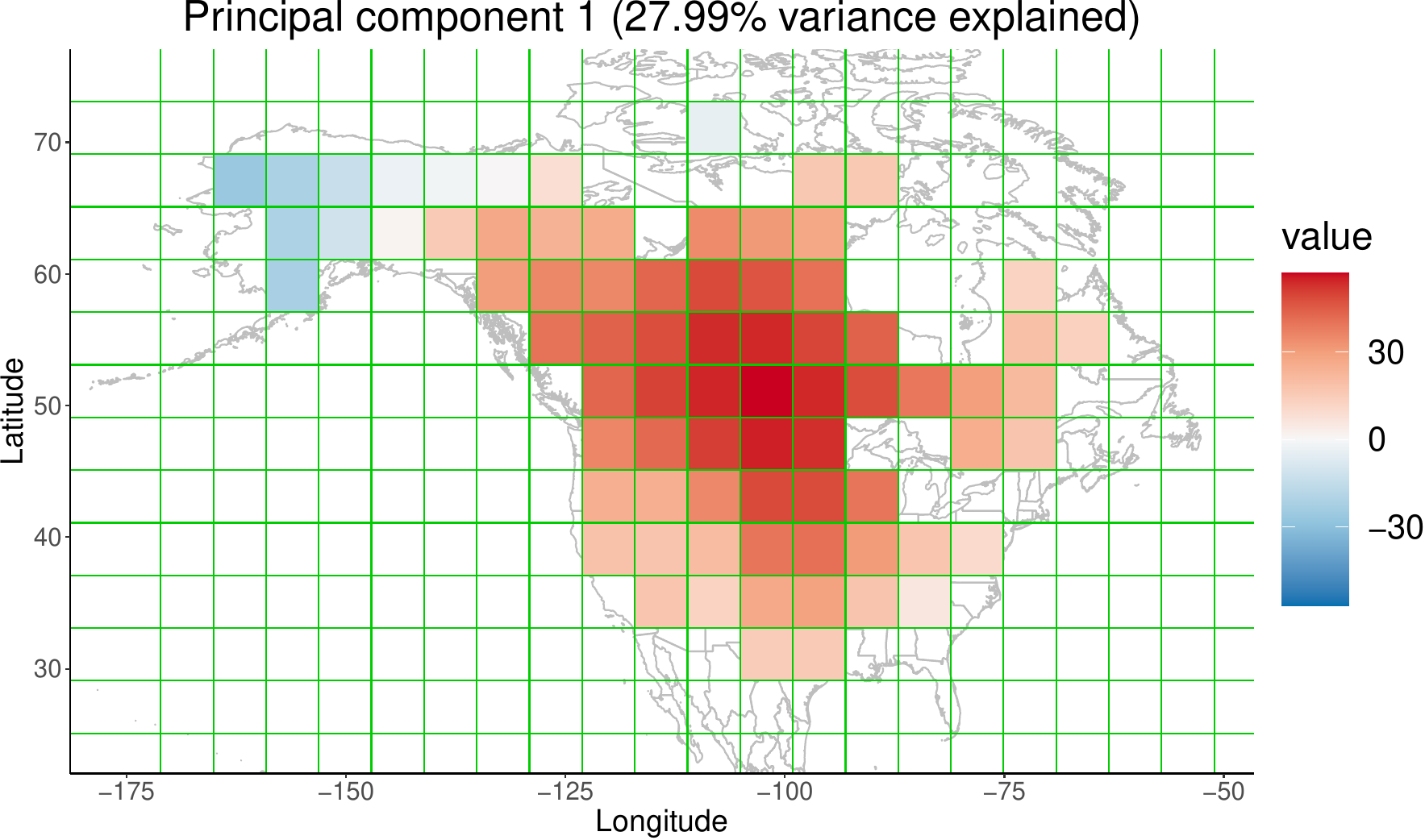}
  \caption{Left panel: cumulative percentage of variability explained by PCA. Right panel: spatial map of the first principal component, where a strong anticorrelated teleconnective pattern is clear.}
  \label{fig:eof}
\end{figure}

We conduct a 10-fold cross-validation comparison to judge how well various probabilistic generative models can reproduce correlation, tail dependence, and spatial extent in the ERA5 temperature data.
First, we separate the set $\{1,\dots,n_{\text{dat}} \}$ into $k$ disjoint sets $\mathcal{I}_1,\dots,\mathcal{I}_k$.
Each generative model listed below is trained on $\mathbf{Y}_{-\mathcal{I}_k}$ for $1 \le k \le 10$, where $\mathbf{Y}_{-\mathcal{I}_k}$ contains all samples of $\mathbf{Y}$ except those from the cross-validation fold $\mathcal{I}_{k}$.
We consider the following models:
\begin{enumerate}
  \item R-vine copula estimated using the \texttt{vinecop} function from \texttt{pyvinecopulib} \citep{pyvinecopulib}. We consider all possible types of bivariate copulas, including their rotations. We also permit nonparametric bivariate copulas based on local likelihood estimators \citep{geenens2017} since the cross-validation performance was noticeably worse upon restricting to only parametric copulas. The model selection procedure from \citet{dissmann2013} automatically chooses the structures of the R-vine, estimates parameters for various bivariate copulas, and then picks the optimal bivariate copula for each vine pair.
  \item Neural Spline Flow \citep{Durkan2019} using \texttt{nflows} \citep{nflows}. See Appendix \ref{app:normalizingflows} for mathematical details.
  \item PCA basis expansion \eqref{eq:basisflow} with Neural Spline Flow for the latent vector. Specifically, we create basis functions from the first $\ell$ eigenvectors of the SVD of the $d \times n_{\text{dat}}$ copula data matrix, where the data is transformed to have standard normal marginals. We show results for $\ell=25$ principal components, which explained 96.4\% of the total variability in the process (see Figure \ref{fig:eof}) and provided a large improvement over $\ell=15$. Appendix \ref{app:comparison} shows how our validation results change as the number of principal components varies.
  \item Generative Moment Matching Network \citep{gmmn} trained with energy distance loss.
\end{enumerate}

\begin{figure}[H]
  \centering
  \includegraphics[scale=.26]{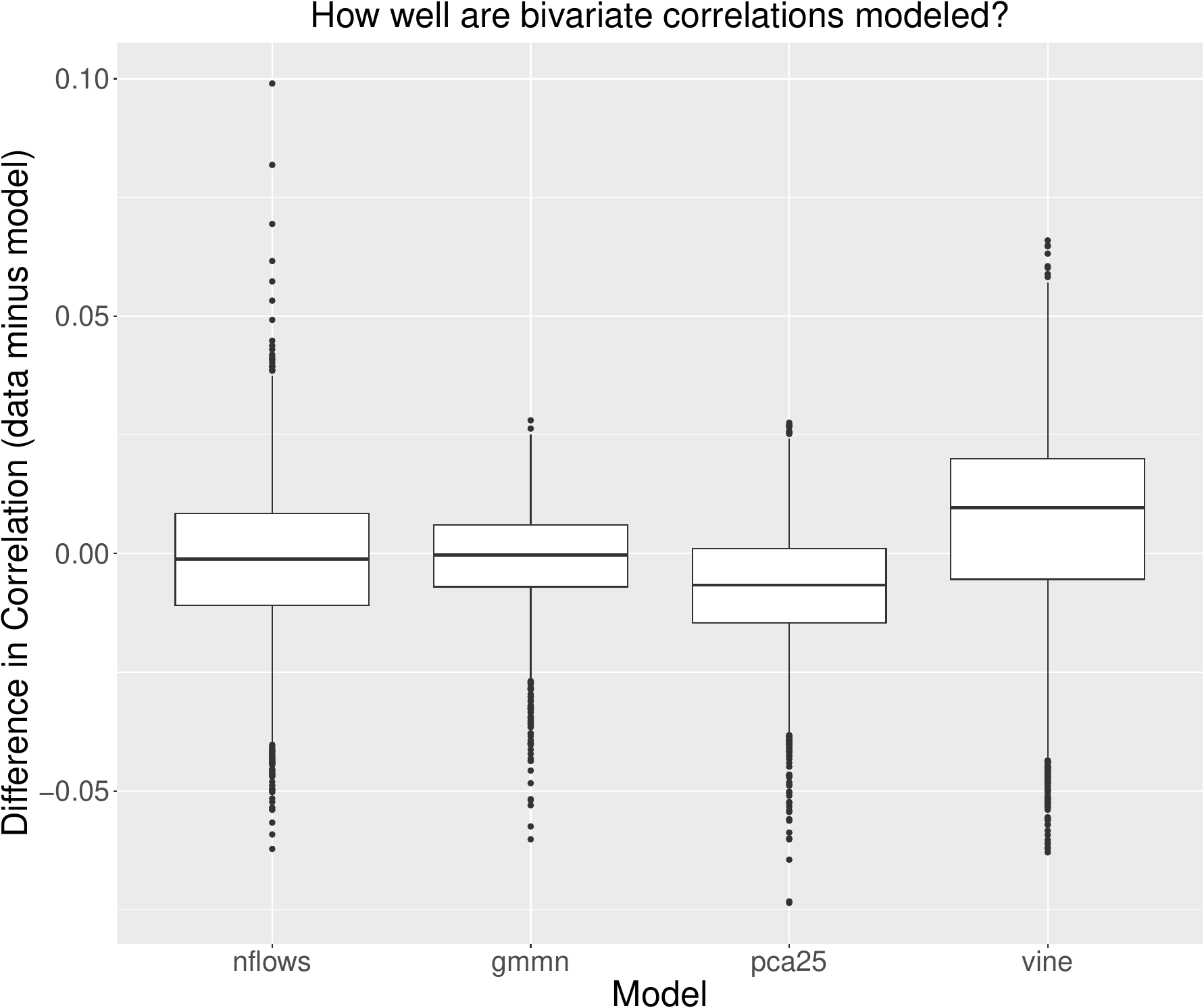}
  \includegraphics[scale=.31]{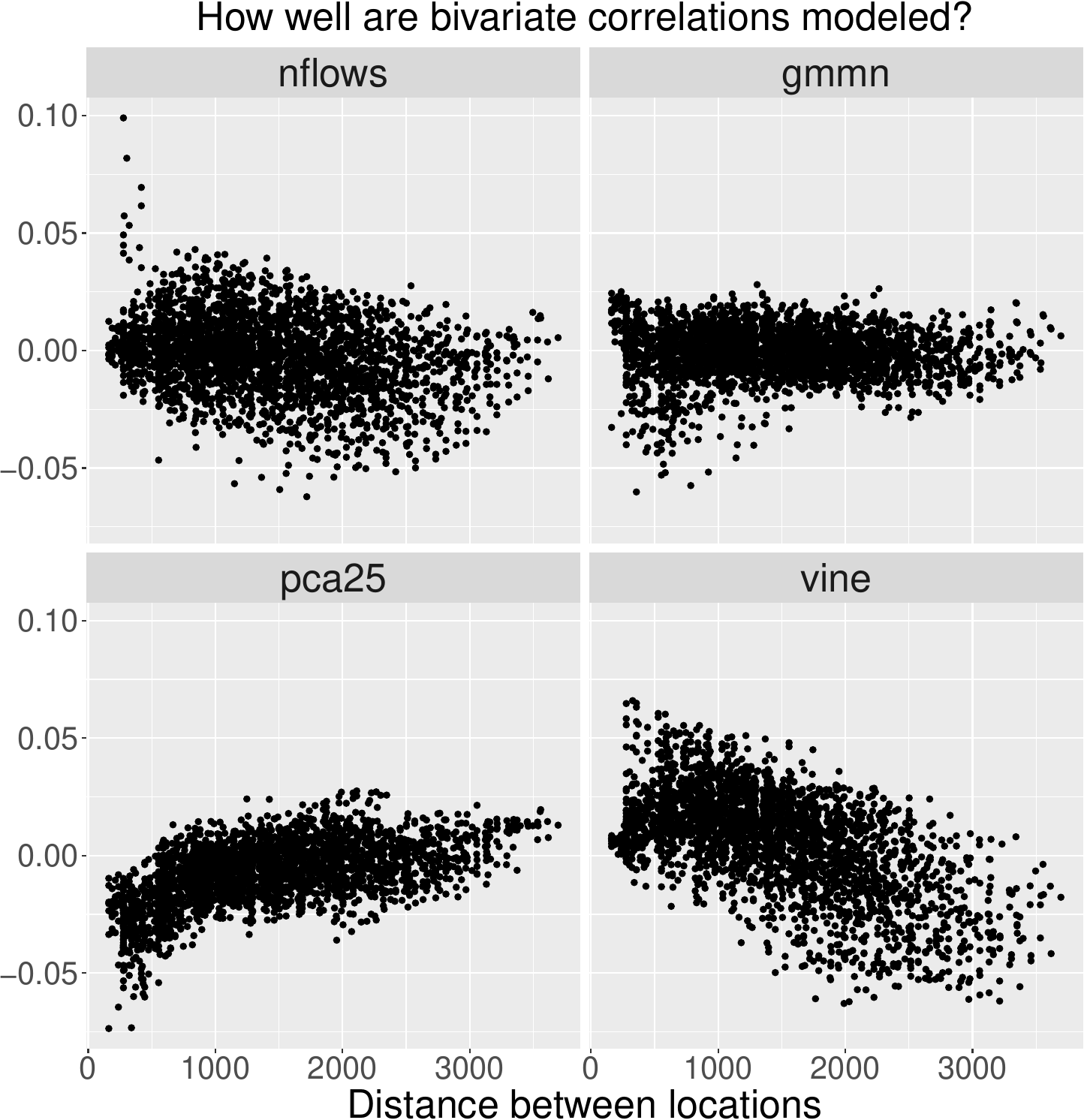}
  \caption{Left panel: boxplots showing the difference in empirical Spearman rank correlation between ERA5 testing data and the probabilistic generative model, averaged over all cross-validation folds. Right panel shows the same statistics on the $y$-axis but plots them as a function of pairwise distance along the $x$-axis.}
  \label{fig:cv_correlation_boxanddist}
\end{figure}

\begin{figure}[H]
  \centering
  \includegraphics[scale=.26]{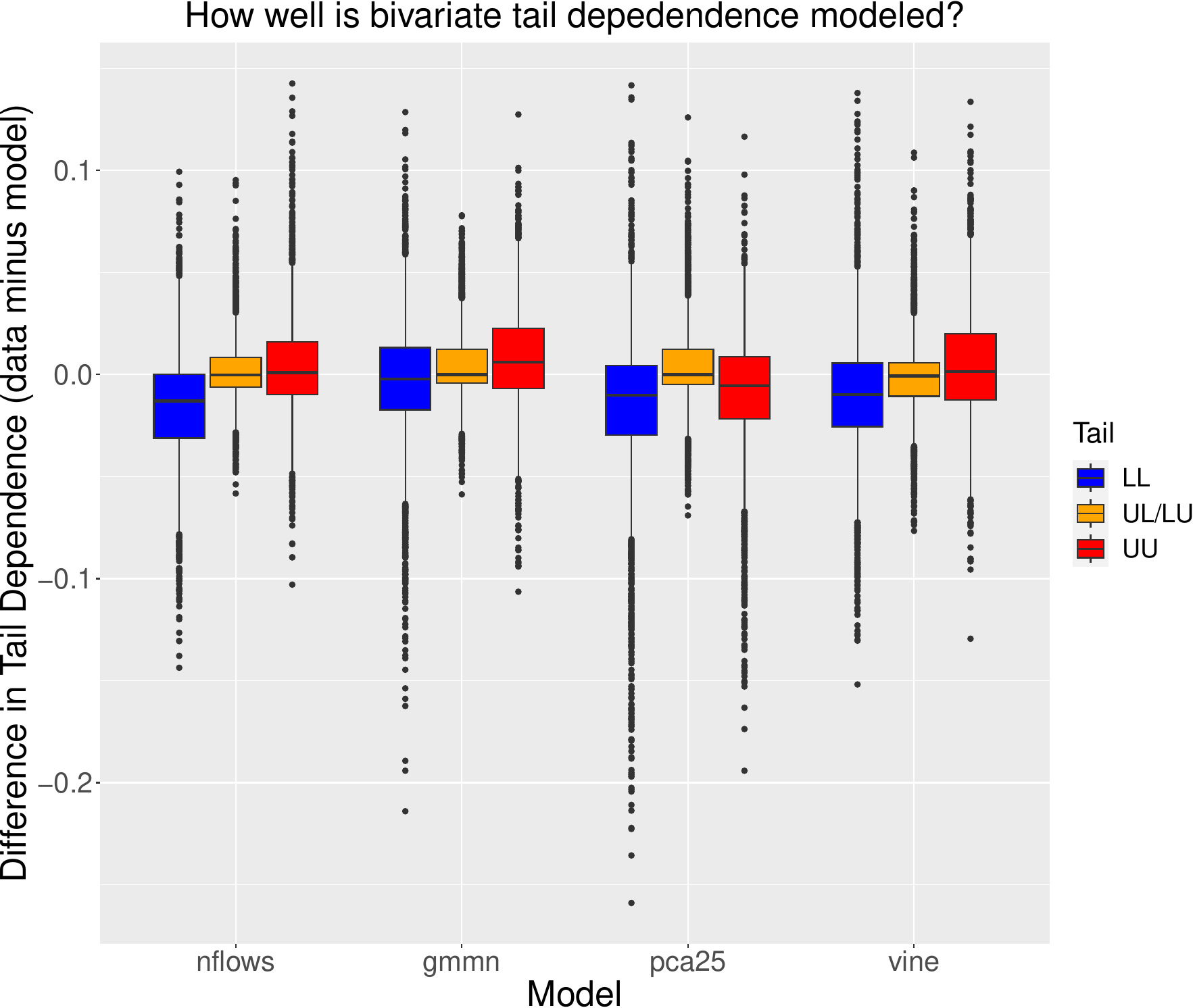}
  \includegraphics[scale=.31]{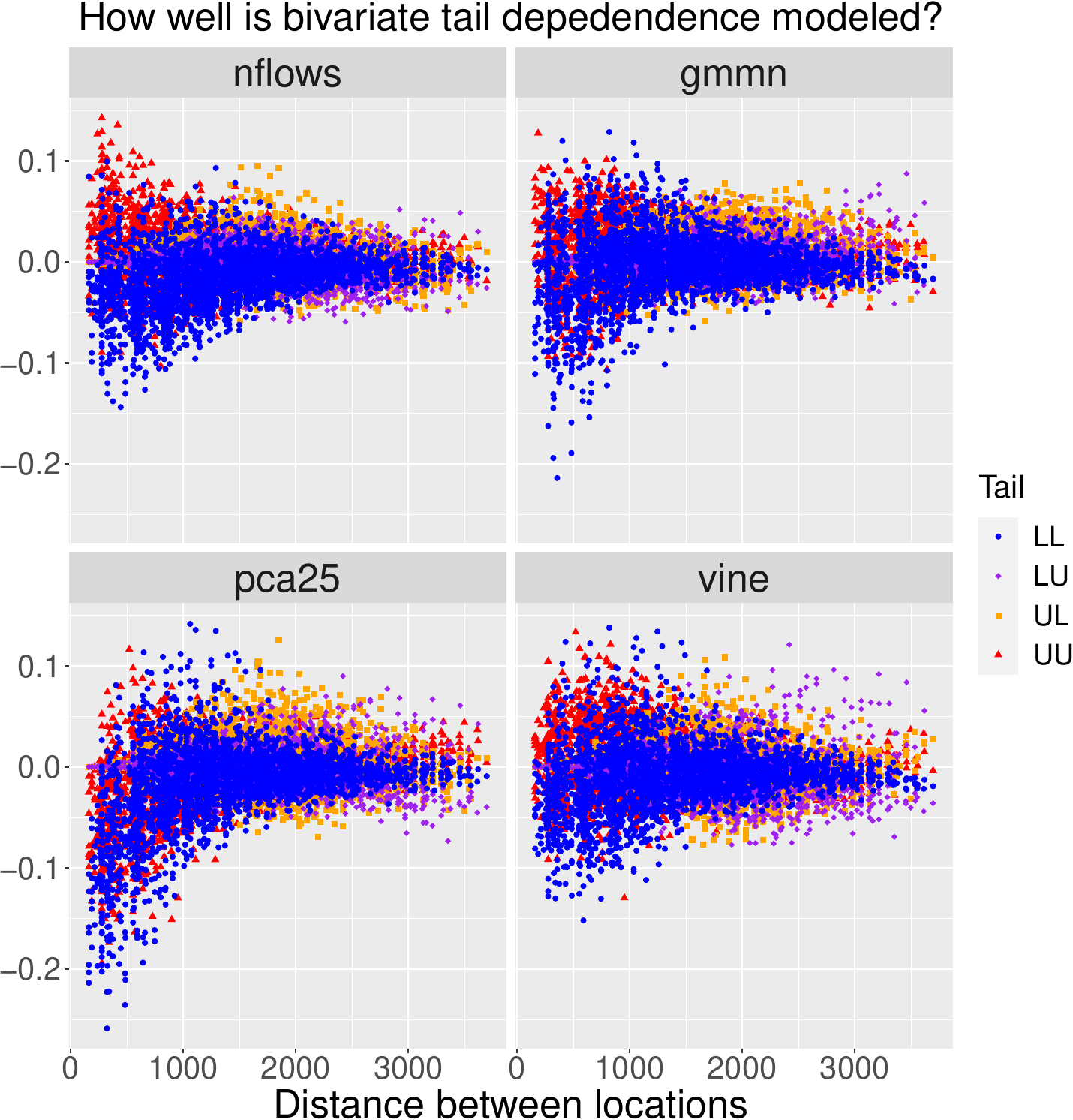}
  \caption{Left panel: boxplots showing the difference in empirical ($u=0.95$) tail dependence between ERA5 testing data and the probabilistic generative model, averaged over all cross-validation folds. Note that $\lambda_{UL}$ and $\lambda_{LU}$ are grouped together in this case. Right panel has the same $y$-axis but plots the values as a function of pairwise distance along the $x$-axis.}
  \label{fig:cv_taildep_boxanddist}
\end{figure}

\begin{figure}[H]
  \centering
  \includegraphics[scale=.26]{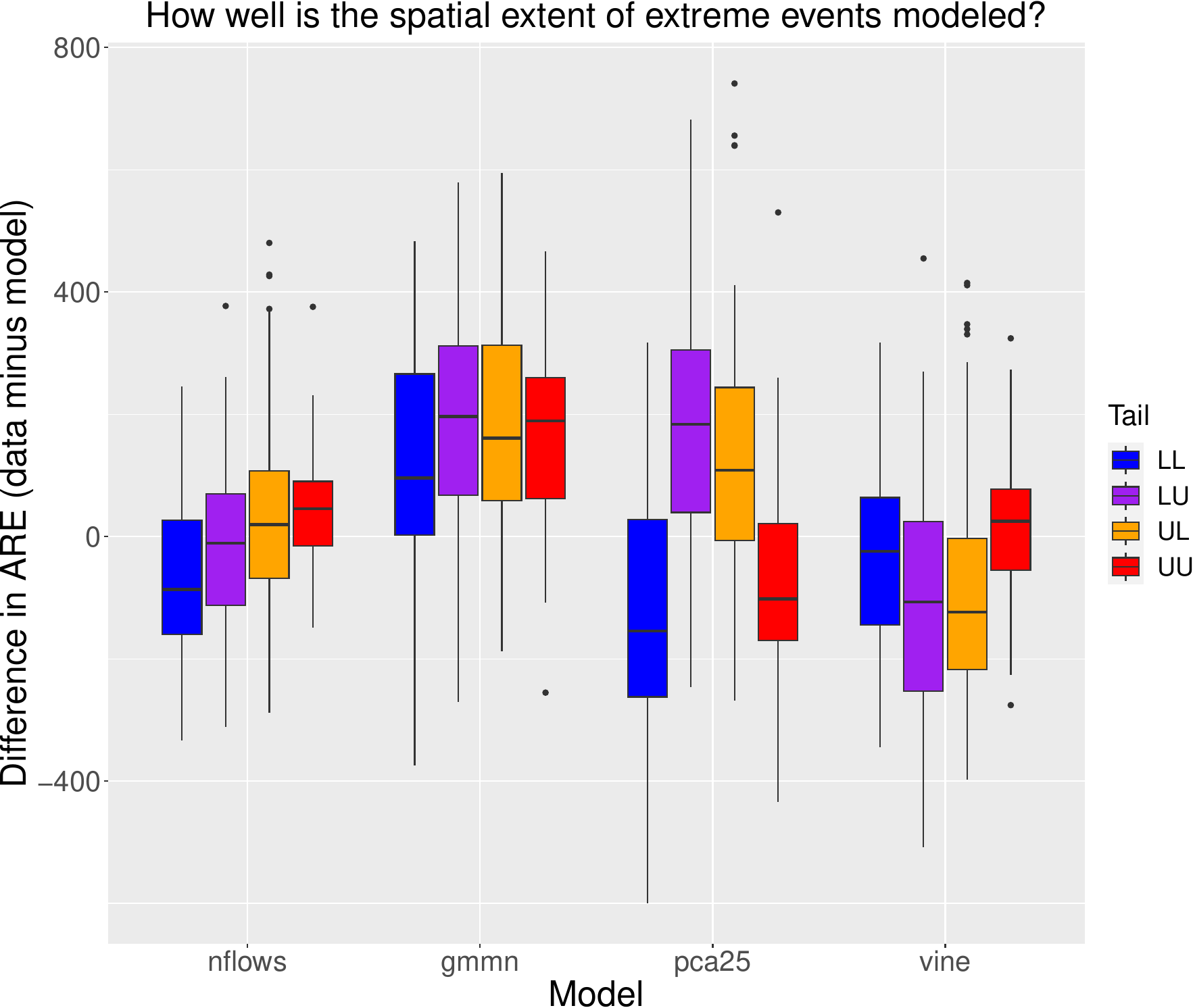}
  \includegraphics[scale=.31]{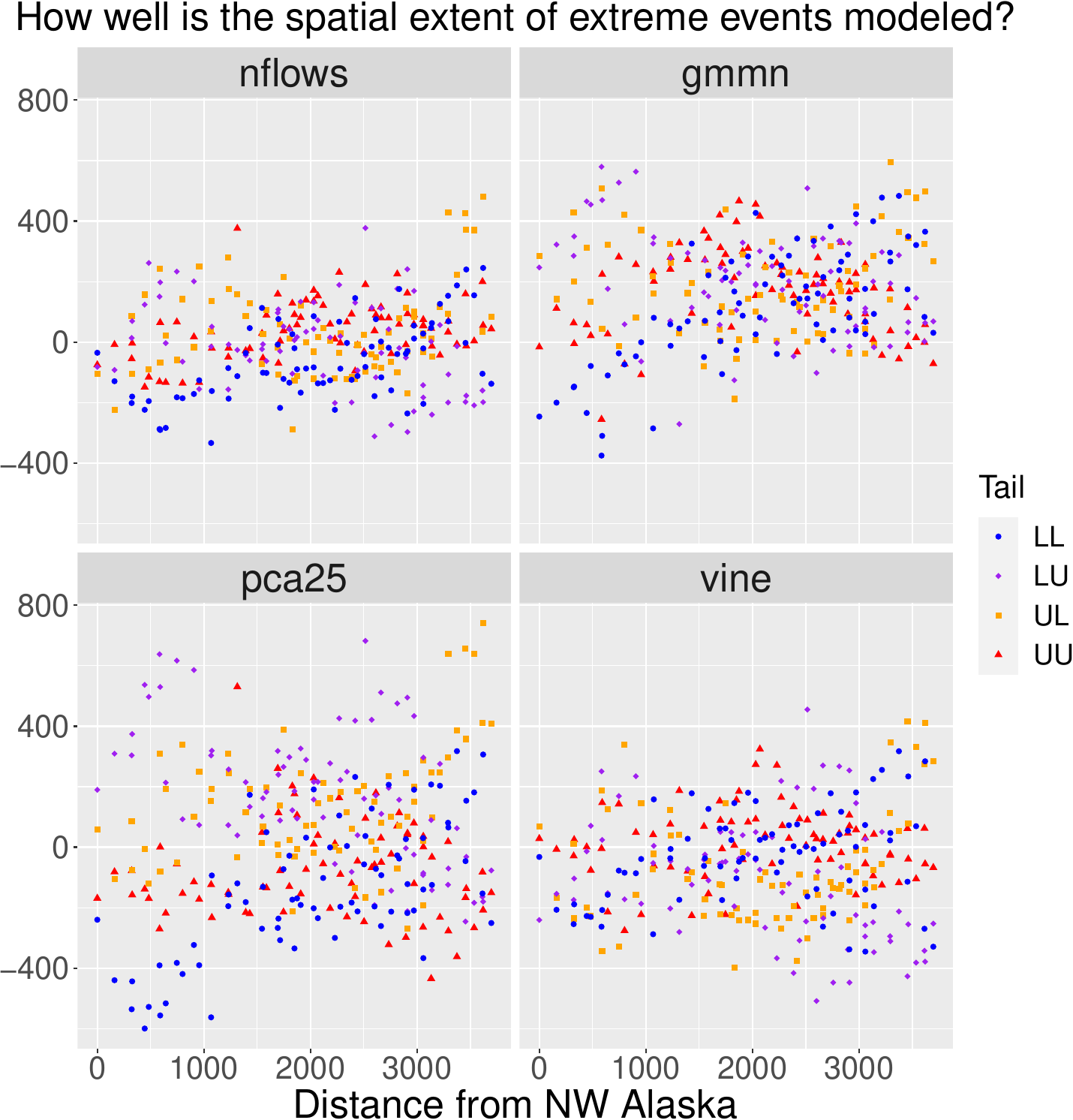}
  \caption{Left panel: boxplots showing the difference in empirical ($u=0.95$) ARE between ERA5 testing data and the probabilistic generative model, averaged over all cross-validation folds. Right panel shows the same statistics on the $y$-axis but plots them as a function of distance along the $x$-axis. In this case, the distance is taken from the reference point in NW Alaska marked with a black dot in Figure \ref{fig:maptaildependencies}.}
  \label{fig:cv_spatialextent_boxanddist}
\end{figure}

Each neural network model has roughly 370,000 parameters, and they are trained for 20,000 epochs using a learning rate of 0.0001 in the ADAM optimizer \citep{KingBa15} and batch size of 100.
Once the training on $\mathbf{Y}_{-\mathcal{I}_k}$ is complete, we generate a new set of $n_{\text{gen}}=10^6$ samples (denoted $\tilde{\mathbf{Y}}_{k}$) in order to evaluate the model fit with respect to the held out data $\mathbf{Y}_{\mathcal{I}_k}$.
For each gridbox $i \in \{1,\dots,d\}$ and $j >i$, we calculate:
\begin{enumerate}
  \item $\frac{1}{10} \sum_{k=1}^{10} \rho_{i,j}(\mathbf{Y}_{\mathcal{I}_k}) - \rho_{i,j}(\tilde{\mathbf{Y}}_k)$, where $\rho_{i,j}(\mathbf{Y}_{\mathcal{I}})$ is the empirical Spearman rank correlation between the $i$th and $j$th gridbox, calculated using realizations $\mathbf{Y}_{\mathcal{I}}$. 
  \item $\frac{1}{10} \sum_{k=1}^{10} \lambda_{UU,i,j}(\mathbf{Y}_{\mathcal{I}_k}) - \lambda_{UU,i,j}(\tilde{\mathbf{Y}}_k)$, where $\lambda_{UU,i,j}(\mathbf{Y}_{\mathcal{I}})$ is the bivariate tail dependence $\lambda_{UU}$ between the $i$th and $j$th gridbox, calculated at the $u=0.95$ quantile using realizations $\mathbf{Y}_{\mathcal{I}}$. This is performed for four tail configurations as in \eqref{eq:empiricaltaildependence}.
  \item $\frac{1}{10} \sum_{k=1}^{10} \alpha_{UU,i}(\mathbf{Y}_{\mathcal{I}_k}) - \alpha_{UU,i}(\tilde{\mathbf{Y}}_k)$, where $\alpha_{UU,i}(\mathbf{Y}_{\mathcal{I}})$ is the empirical ARE conditional $\alpha_{UU}$ on gridbox $i$ being extreme, calculated at the $u=0.95$ quantile using realizations $\mathbf{Y}_{\mathcal{I}}$. This is performed for four tail configurations as in \eqref{eq:spatialextent}. 
\end{enumerate}
Note that these values are calculated with respect to the uniform copula, so if the testing data or simulations correspond to standard normal marginals, they are transformed to uniformity via the standard normal cdf.
There are $d$ differences in Step 3 and $d(d-1)/2$ in Steps 1 and 2 that are used to evaluate and visualize the cross-validated model performance.
A simple solution is to average these values and display them in boxplots grouped by model.
This provides a quick summary of model performance but collapses information over space.
Instead, we can plot the values as a function of distance. 
For the pairwise statistics in Steps 1 and 2, there is a natural notion of distance, but each statistic in Step 3 is associated with a single gridbox.
Therefore, we plot the spatial extent as a function of distance from an arbitrary location, which is chosen to be the gridbox in NW Alaska marked with a black dot in Figure \ref{fig:maptaildependencies}.
Figures \ref{fig:cv_correlation_boxanddist}, \ref{fig:cv_taildep_boxanddist}, and \ref{fig:cv_spatialextent_boxanddist} show these two types of summary plots for correlation, tail dependence, and spatial extent, respectively.

In terms of spatial correlation, \texttt{gmmn} performs best, which is unsurprising since the generator is trained to match all moments of the distribution.
Meanwhile, \texttt{nflows} tends to underestimate local positive correlations compared to the data distribution; this bias disappears when the flows are combined with principal components, likely because the basis functions add a notion of spatial continuity to the model.
The PCA model noticeably overestimates $\alpha_{UU}, \alpha_{LL}, \lambda_{UU}$, and $\lambda_{LL}$, which measure dependence along the main diagonal of the distribution.
Adding more basis functions does help correct these biases (see Figure \ref{fig:correlation_boxplot_nocv}).
Even so, the PCA model remains competitive with the other models that do not perform any dimension reduction.
Similarly, \texttt{gmmn} also overestimates $\lambda_{UU}$ and $\lambda_{LL}$ at nearby locations, but interestingly it underestimates $\alpha_{UU}$ and $\alpha_{LL}$ along with the other two types of spatial extent.
Overall, \texttt{nflows} compares favorably with \texttt{gmmn}, performing slightly better for tail statistics and slightly worse for bulk statistics. 

The vine copula is also competitive with the deep learning methods despite its simple formulation.
Figure \ref{fig:cv_correlation_boxanddist} illustrates a weakness of the vine copula in that it markedly underestimates the probability of teleconnected extremes where NW Alaska is abnormally warm and western U.S.\ is abnormally cold.
Meanwhile, \texttt{nflows} models this type of dependence relatively well.
Since both models have tractable densities, we can compare the cross-validated loglikelihood
$
\frac{1}{10} \sum_{k=1}^{10} \ell(\mathbf{Y}_{\mathcal{I}_k}; \mathbf{Y}_{-\mathcal{I}_k})
$
where $\ell(\mathbf{Y}_{\mathcal{I}_k}; \mathbf{Y}_{-\mathcal{I}_k})$ is the loglikelihood of a model trained on $\mathbf{Y}_{-\mathcal{I}_k}$, evaluated with the testing data $\mathbf{Y}_{\mathcal{I}_k}$.
The cross-validated loglikelihoods for \texttt{nflows} and the vine copula are 34081.21 and 31017.97, respectively.
Interestingly, the parametric vine copula produces a higher cross-validated loglikelihood of 33990.49, but its performance in other cross-validation metrics were worse than the nonparametric vine copula.
This data analysis provides an example where moving from traditional statistical methodology vine copulas to normalizing flows provides better performance in modeling some aspects of extremes.

\begin{figure}[H]
  \centering
  \includegraphics[scale=.94]{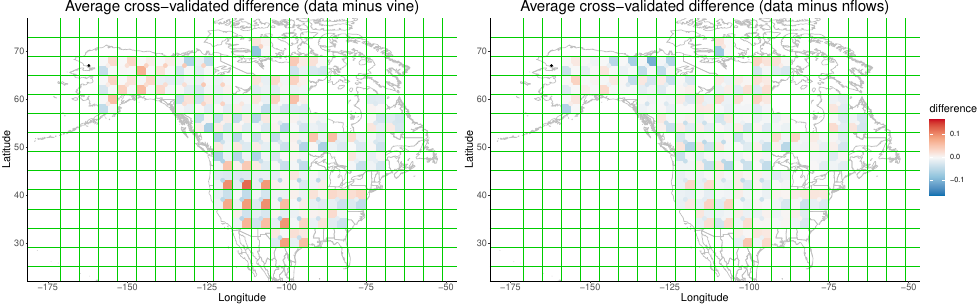}
  \caption{Displaying results from the right panel of Figure \ref{fig:cv_taildep_boxanddist} in a spatial map.}
  \label{fig:cv_fivestat_distance}
\end{figure}


\section{Discussion and Conclusions}
\label{sec:discussion}

Teleconnected patterns where distant locations experience oppositely-signed temperature extremes motivate the need for multivariate extreme models that consider the entire distribution.
This work is an initial step towards spatial modeling of opposite-tail extremes, primarily addressing the asymmetric patterns in the spatial distribution of surface air temperature.
We propose a copula model with Bulk-And-Tails marginal distributions whose dependence structure is controlled by a normalizing flow.
Overall, we found the model was easy to implement and reproduced many complicated asymmetric spatial patterns in winter temperatures in North America, such as anticorrelated teleconnecive patterns in the bulk and tails of the distribution.
Designing models for teleconnected spatial extremes can help create resilience against severe weather events caused by atmospheric blocking.

We also considered modeling dependence with a normalizing flow as the stochastic weights of a principal component basis expansion.
Ordering the principal components by variance provides a natural solution to the autoregressive structure in the normalizing flow.
An issue raised by \citet{jiang2020} is that principal components are more suitable for describing the bulk of the distribution rather than the joint tails, so it may be worth exploring basis representations, including nonlinear ones \citep{Tagasovska2019}.
Principal components add a notion of spatial smoothness that normalizing flows themselves lack, as seen in the biased local correlations in Figure \ref{fig:cv_correlation_boxanddist}.
Convolutional layers are useful when trying to model spatial dependence with neural networks, but they require gridded data, which is not available in our case.
Moreover, convolutions are not invertible operators, complicating their use with normalizing flows; \citet{karami2019} construct invertible convolutions for this purpose.

Flows are more expressive than vine copulas, a standard statistical tool for dependence modeling.
This greater flexibility comes with a cost: the flow is only a copula up to an approximation of the uniform distribution, and analytic expressions of tail dependence coefficients are unavailable.
The practical implications of the first point are that the marginals estimated in the first step of our copula model will not be exact. 
We are unaware of any deep learning models that can overcome this limitation.
As mentioned in Section \ref{sec:related}, this bias can be partially mitigated with marginal corrections.
With vine copulas, the analytic expression for tail dependence involves recursive numerical computation of multivariate integrals, so tractability is not necessarily better than resorting to empirical estimation of tail dependence coefficients.
A common simplifying assumption in the vine copula literature is that the parameters of the conditional copulas are constant with respect to the conditional variables.
While there are efforts to construct vine copulas without this limitation \citep{acar2012,stober2013,zhang2018}, this type of parameter-level conditional dependence is a natural byproduct of normalizing flows.
Balancing the expressiveness of neural networks with the interpretability of classical parametric statistical models is a growing focus of spatial deep learning \citep{wikle2023}.


The biggest issue with our model from a spatial extremes point-of-view is that it lacks parameters to control the behavior of the joint tails of the distribution.
Except in trivial cases, it is impossible to calculate tail dependence coefficients of a flow model.
With other probabilistic generative models that lack an explicit form for their probability density, the task of parameterizing joint tail behavior seems even more difficult.
Indeed, the inability of neural networks to describe out-of-sample events is a prominent obstacle in the deep learning community \citep{Nalisnick2019, Kirichenko2020}.
In this work, we have demonstrated the effectiveness of several nonparametric models in reproducing asymmetries of the (in-sample) spatial distribution of surface air temperature, which is a necessary first step towards a spatial process model that is suitable for opposite-tail teleconnective extremal dependence.

\clearpage
\acknowledgments {Mitchell Krock and Julie Bessac acknowledge support from the U.S. Department of Energy, Office of Science, Office of Advanced Scientific Computing Research (ASCR), Contract DE-AC02-06CH11357.
Thanks to Adam Monahan for useful discussions.}


\datastatement {ERA5 data is publicly available at \url{https://cds.climate.copernicus.eu/cdsapp#!/dataset/reanalysis-era5-single-levels?tab=overview}.}

\newpage

\appendix[A]

\section{Further into ERA5 tails}
\label{app:farthertails}
\begin{figure}[H]
    \centering
    \includegraphics[scale=.25]{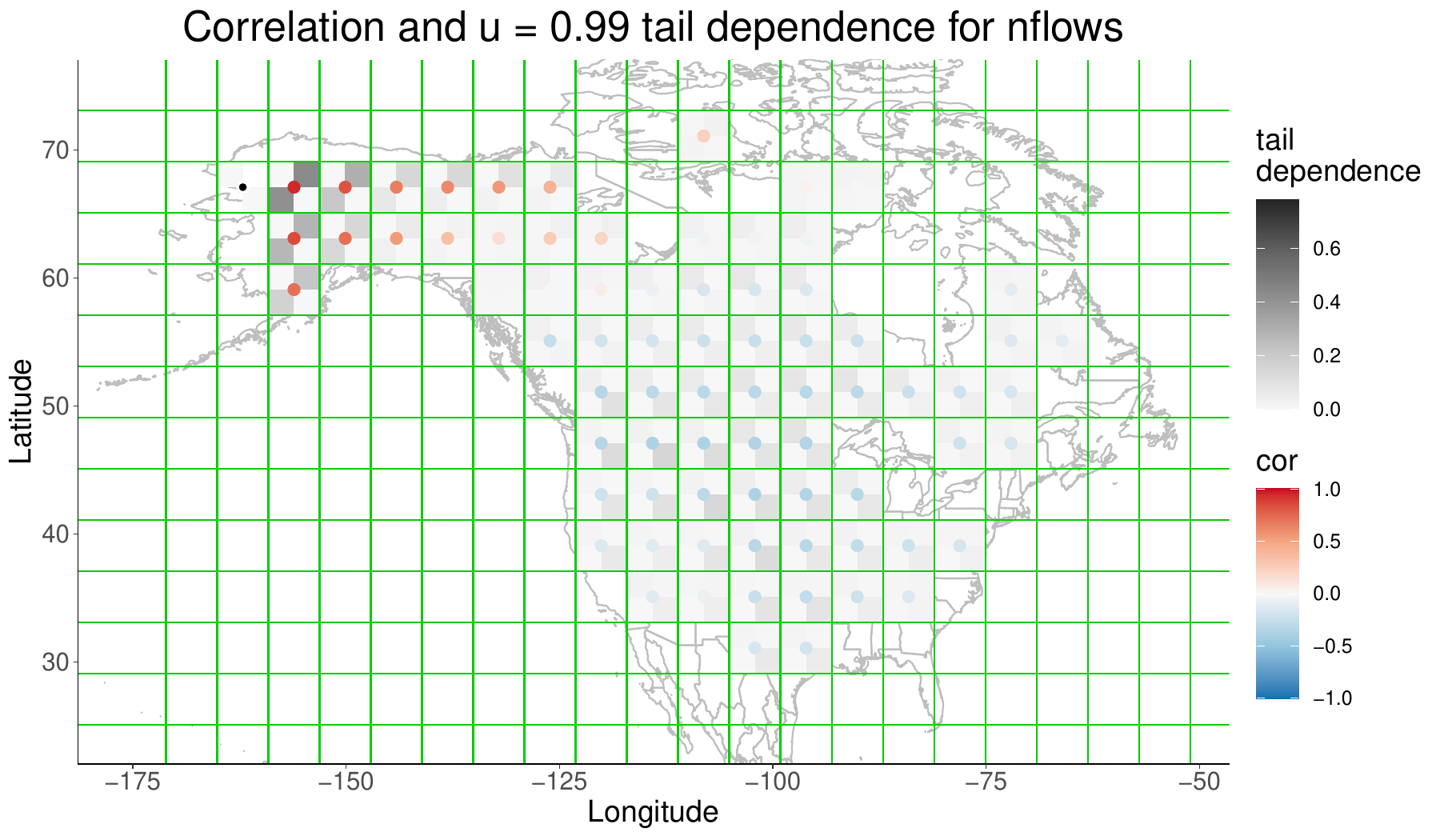}
    \includegraphics[scale=.25]{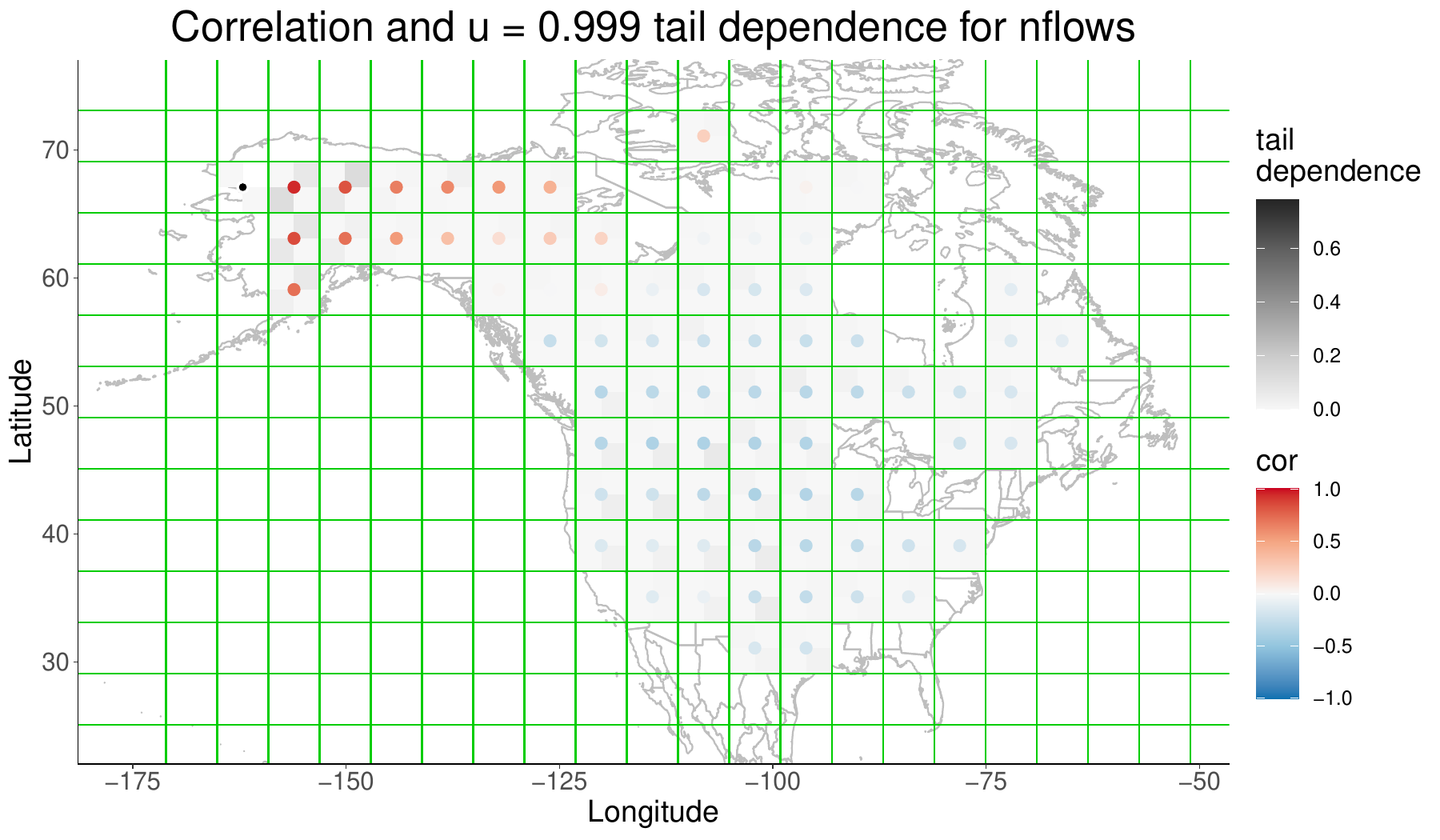}
    \includegraphics[scale=.25]{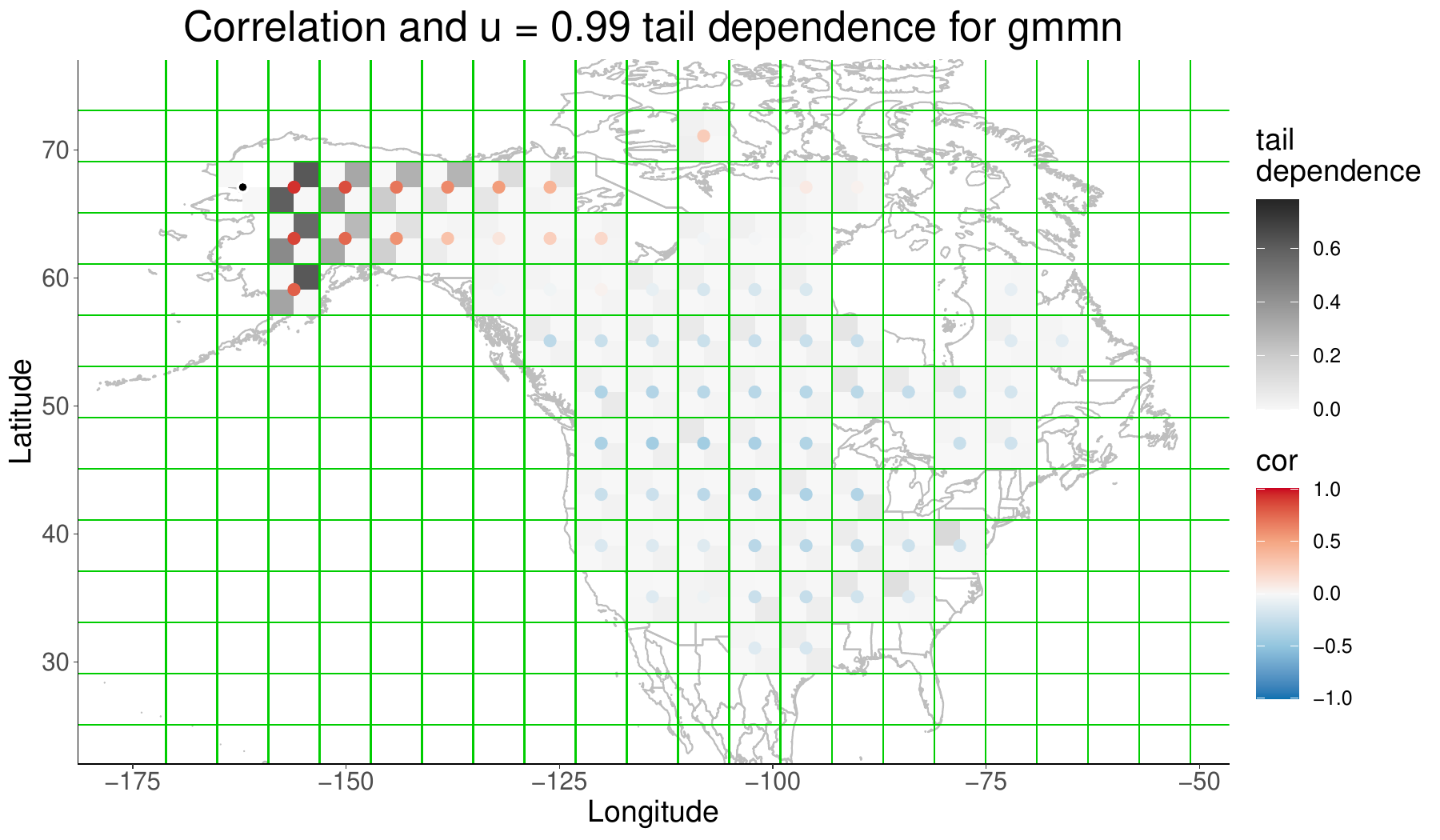}
    \includegraphics[scale=.25]{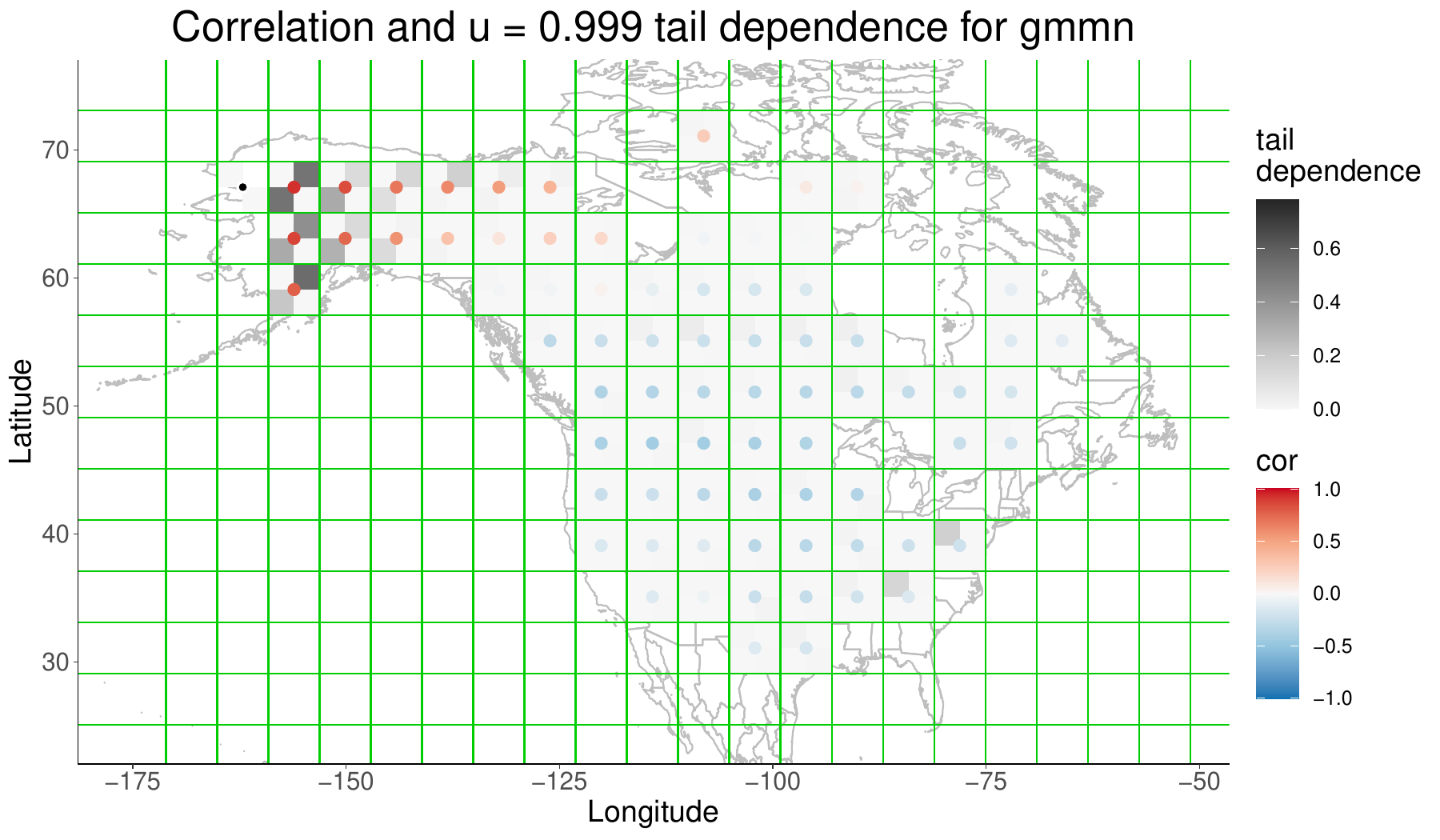}
    \includegraphics[scale=.25]{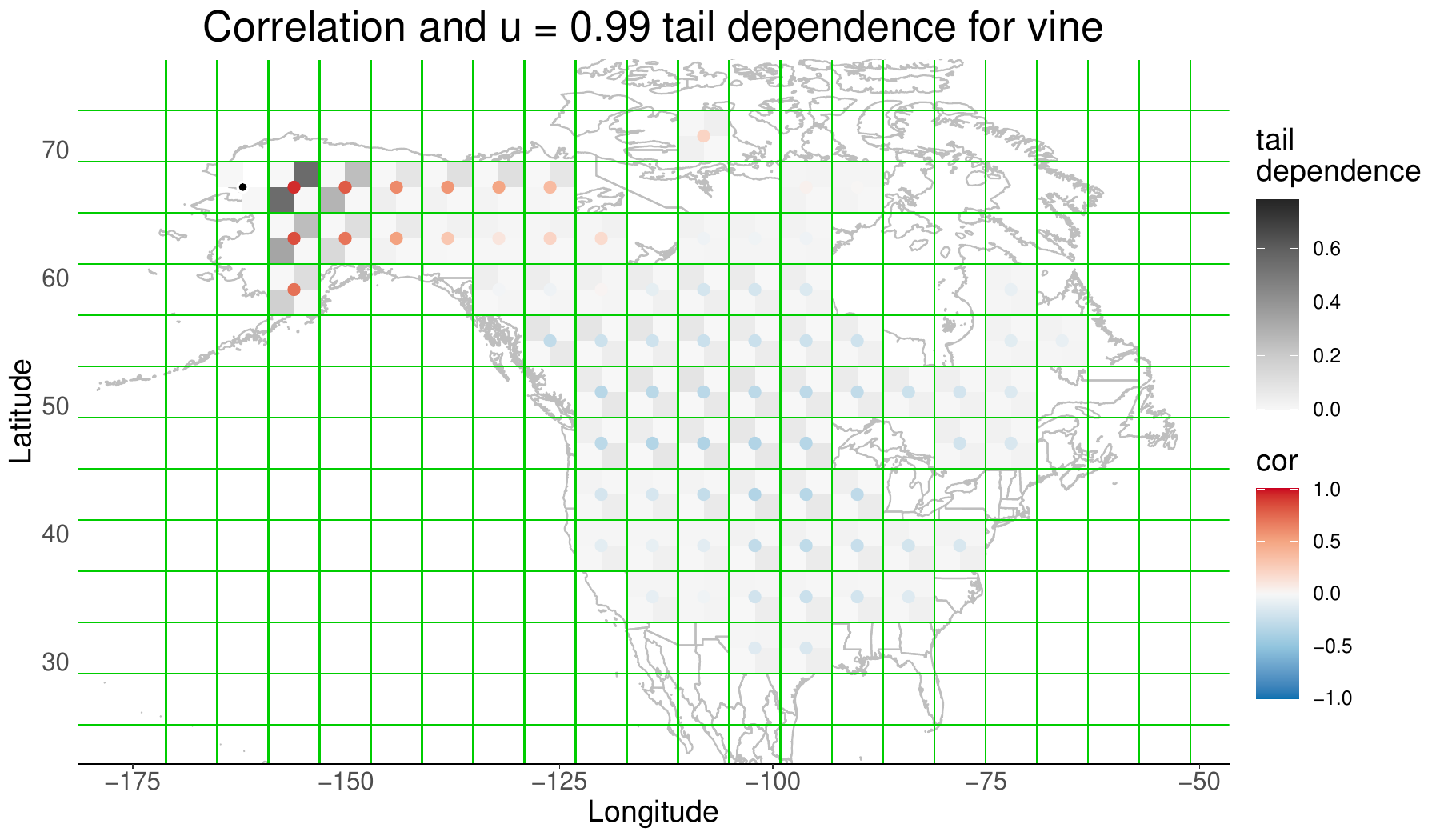}
    \includegraphics[scale=.25]{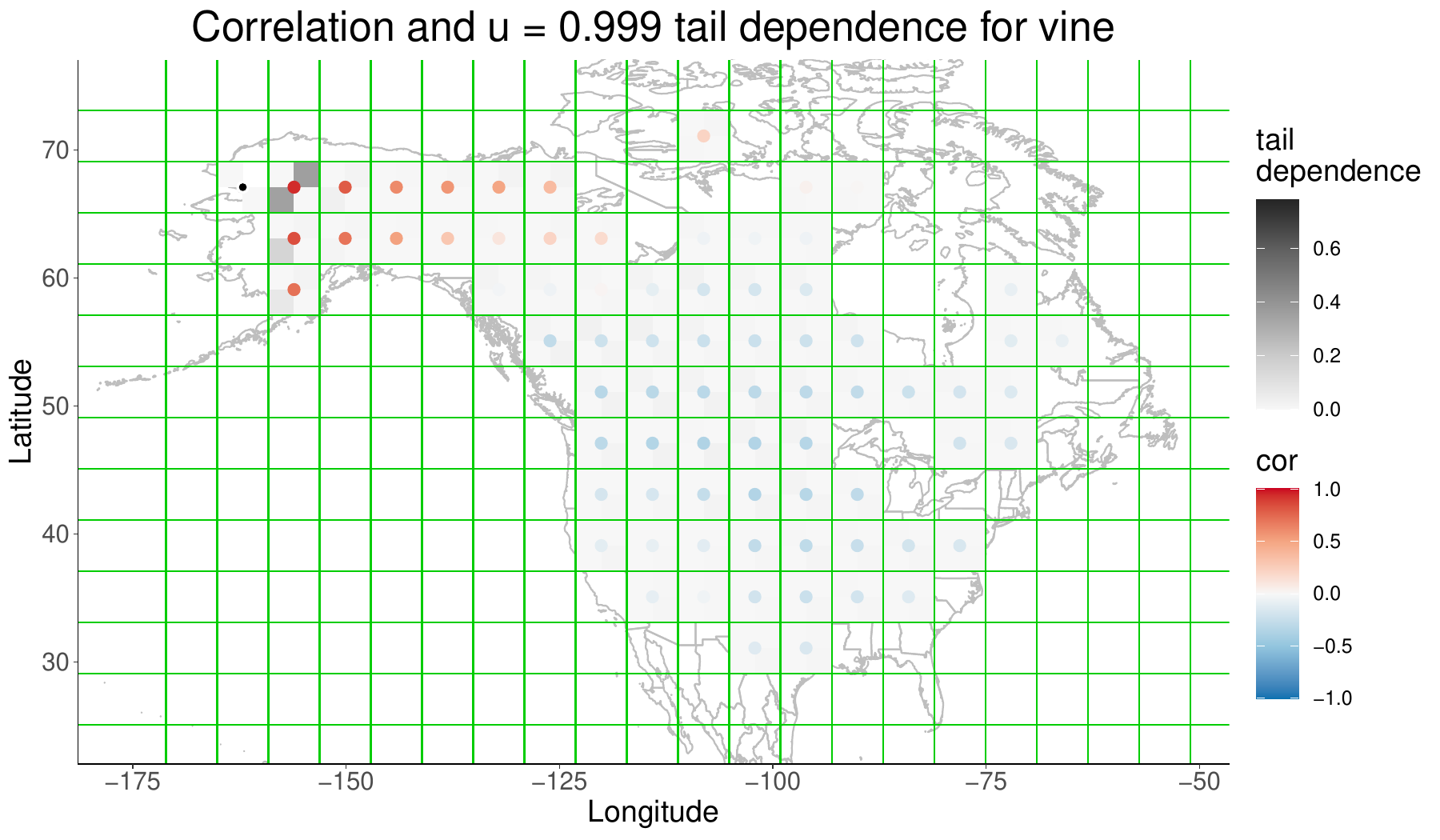}
    \caption{
      Displaying the estimated tail dependence coefficients at quantiles $u = 0.99, 0.999$, calculated using $10^6$ samples from the generative model.
    }
    \label{fig:u99999}
  \end{figure}

In Figure \ref{fig:u99999}, we show the tail dependence of our probabilistic generative models beyond the $u=0.95$ quantile.
Values are again based on $n_{\text{gen}} = 10^6$ unconditional samples from the model (without any cross-validation).
Clearly, the models in their current form do not exhibit long-range asymptotic dependence as $u \uparrow 1$.

\section{Normalizing Flow}
\label{app:normalizingflows}
Here we provide additional details about the parametric form of the normalizing flow $g_\theta : \mathbb{R}^\ell \to \mathbb{R}^\ell$.
Recall that $g_\theta(\cdot)$ must be an invertible function with tractable Jacobian determinant.
Flows are often constructed in an autoregressive fashion that takes the noise vector $\mathbf{Z} \in \mathbb{R}^\ell$ and outputs the $i^{\mathrm{th}}$ component of $\mathbf{Y} = g_\theta(\mathbf{Z})$ as $Y_i = h(Z_i; c_i(\mathbf{Z}_{<i}))$, where $h(\cdot;\theta) : \mathbb{R} \to \mathbb{R}$ is a monotonically increasing function whose parameters $\theta$ are obtained as the output of a neural network $c_i(\cdot)$ which takes $\mathbf{Z}_{<i} = (Z_1,\dots,Z_{i-1})^{\mathrm{T}}$ as input.
In practice, $c_i(\mathbf{Z}_{<i})$ for $i=1,\dots,\ell$ are not the output of $\ell$ separate neural networks, but rather the output of a single pass through a neural network that is appropriately ``masked'' to preserve the autoregressive structure \citep{maf}.

By construction, the Jacobian matrix of this autoregressive transformation is triangular and therefore has a trivial determinant.
Moreover, $g_\theta^{-1}(\cdot)$ is readily computed since $h(\cdot;\theta)$ is a bijection; i.e.,\ $Z_i = h^{-1}(Y_i; c_i(\mathbf{Z}_{<i}))$.
However, calculating this inverse is slow since it requires sequential computation in that $c_i(\mathbf{Z}_{<i})$ must be computed before $Z_i$.
This typically means that sampling from an autoregressive flow is $\ell$ times slower than evaluating its likelihood, although in this paper we have used the convention that $g_\theta(\cdot)$ maps from the noise distribution to the data distribution, while most flows are formulated in the opposite direction.

Early flow models consider simple forms for $h(\cdot;\theta)$ (e.g.,\ an affine transformation), limiting their expressive power.
In the Neural Spline Flow, \citet{Durkan2019} define $h(\cdot;\theta)$ with rational quadratic splines, improving upon their previous work that uses cubic splines.
Specifically, they use monotonically-increasing piecewise rational quadratic splines that bijectively maps $[0,1]$ to $[0,1]$.
\citet{Durkan2019} suggest augmenting this mapping with the identity transformation outside $[0,1]$ so that the flow can take unbounded inputs.
We found that mapping from $\mathbb{R}^\ell$ to $[0,1]^\ell$ with the sigmoid function, modeling on $[0,1]^\ell$ with autoregressive rational spline transformations, and then transforming back to $\mathbb{R}^\ell$ with the logit function was a more effective way to handle unbounded inputs than using linear tails.
Note that strictly-increasing componentwise transformations like sigmoid and logit activation functions do not change the copula \citep{Joe14}.
\citet{laszkiewicz2022} observed that using linear tails outside $[0,1]$ leads to the same type of restrictive tail behavior seen in triangular affine flows \citep{jaini2020} as mentioned in Section \ref{sec:statbackground}\ref{sec:marginals}.

\section{Neural Network Architecture}
Here we briefly describe the neural network architectures used in Section \ref{sec:era5} to model the ERA5 temperature data.
These models are based on the \texttt{PyTorch} backend and assume marginal distributions of the pseudo-copula are standard normal rather than standard uniform.

The Generative Moment Matching Network is a standard multilayer perceptron with the following structure:
\begin{verbatim}
  Sequential(Linear(76, 100), ReLU(), Linear(100, 200), ReLU(),
    Linear(200, 400), ReLU(), Linear(400, 400), ReLU(),
    Linear(400, 200), ReLU(), Linear(200, 100), ReLU(),
    Linear(100, 76))
\end{verbatim}

The architecture for the normalizing flow in dimension \texttt{L} is as follows:

\begin{verbatim}
base_dist = distributions.normal.StandardNormal(shape=[L]) 
num_layers = 4
transforms = []
transforms.append(Sigmoid())
for _ in range(num_layers):
  transforms.append(
    MaskedPiecewiseRationalQuadraticAutoregressiveTransform(features=L,
    hidden_features=32))
  transforms.append(RandomPermutation(features=L))
transforms.append(Logit())
\end{verbatim}

\section{More model comparisons}
\label{app:comparison}
We ruled out several models before cross-validation study in Section \ref{sec:era5} due to poor performance.
First, we disregarded GAN methods due to training difficulties.
In particular, we tried two models trained with adversarial loss:  GAN \citep{gan} and Wasserstein GAN \citep{wgan}. 
The Wasserstein GAN uses a different loss function to avoid issues (e.g.,\ mode collapse, vanishing gradients) that are commonly encountered with the original GAN loss function in \citet{gan}.
However, both methods performed poorly on the ERA5 data.

\begin{figure}[H]
  \centering
  \includegraphics[scale=.4]{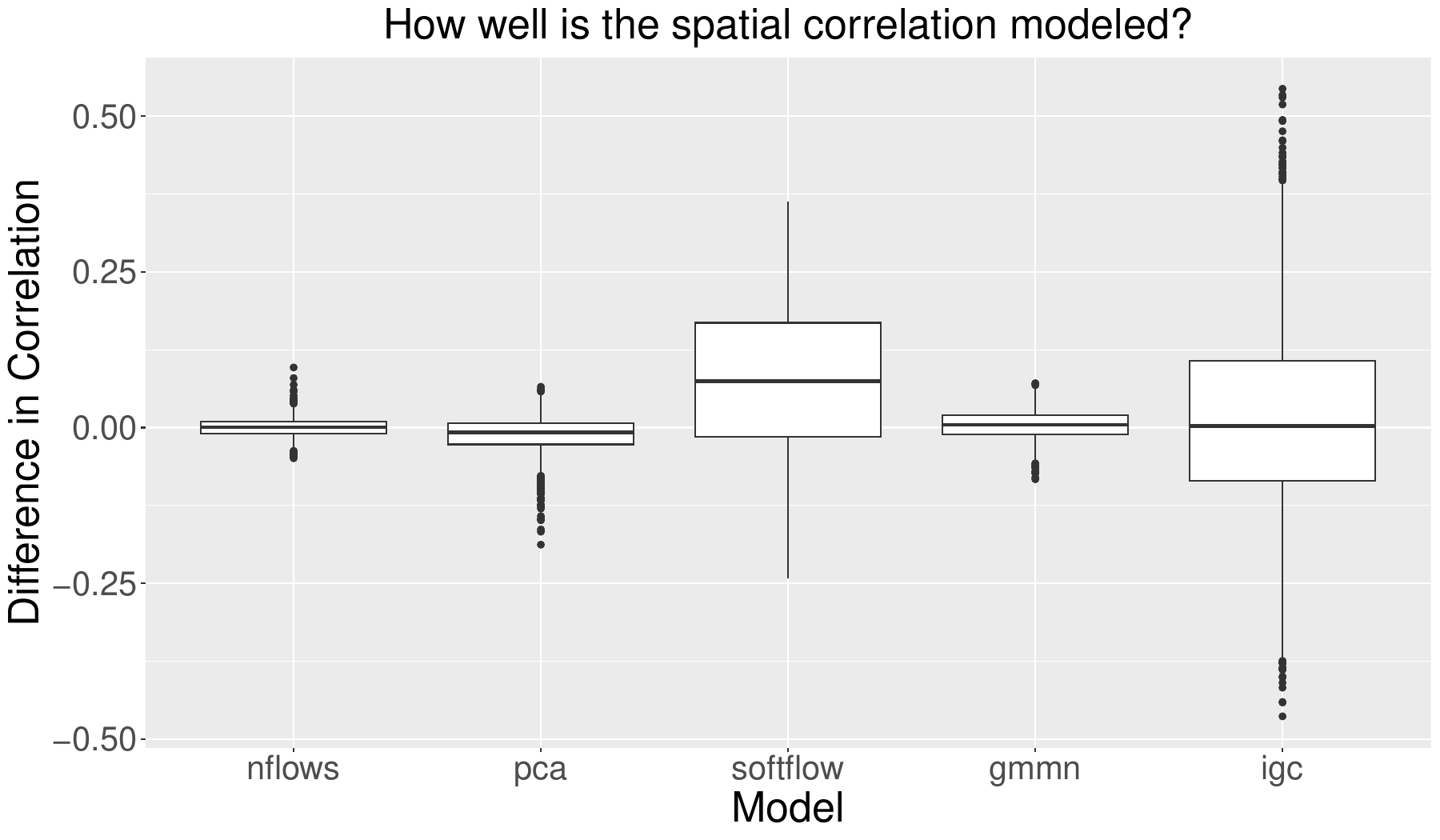}
  \caption{Boxplots showing the difference in empirical Spearman rank correlation between ERA5 data and the probabilistic generative model.}
  \label{fig:correlation_boxplot_nocv}
\end{figure}

\begin{figure}[H]
  \centering
  \includegraphics[scale=.45]{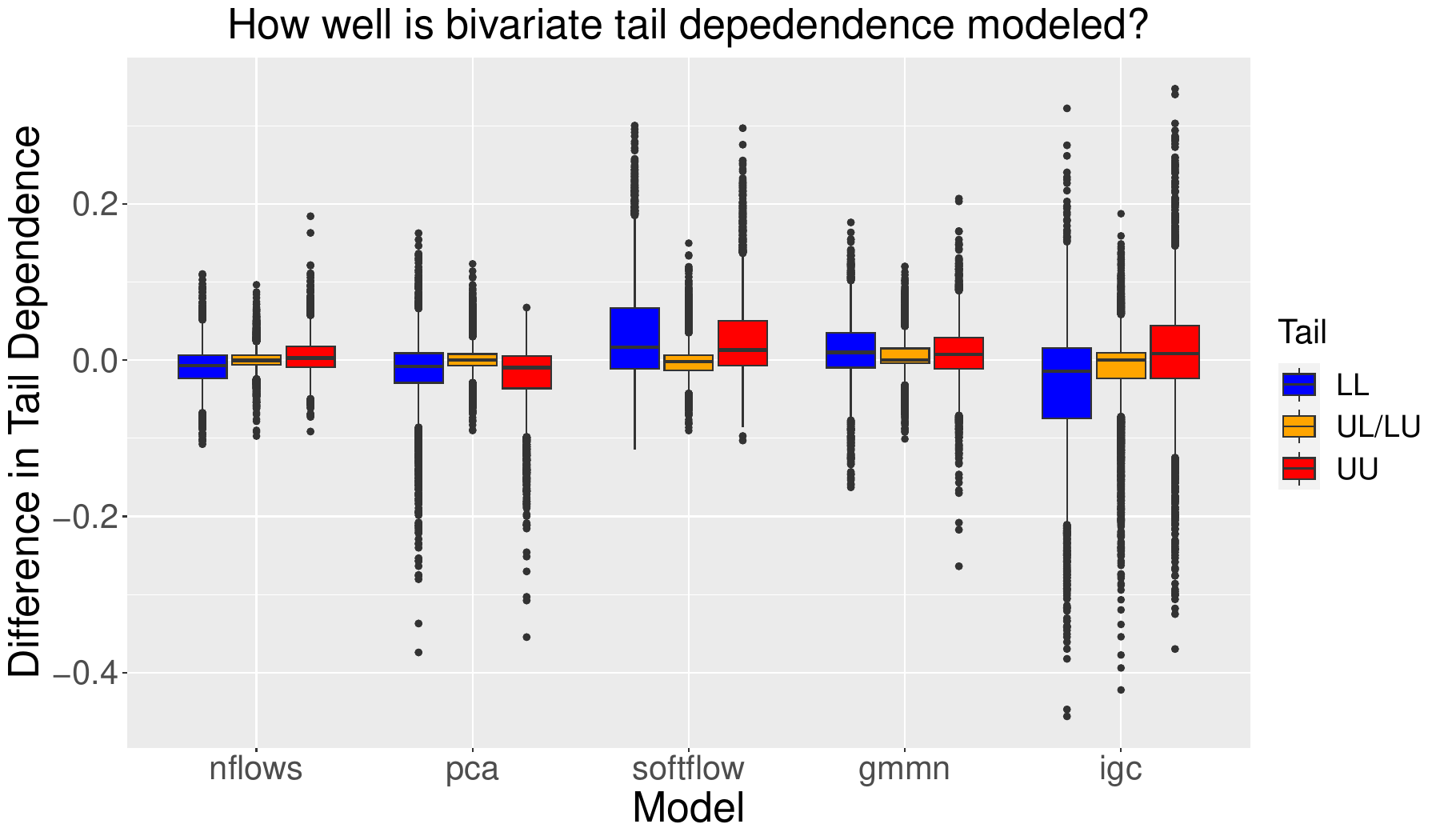}
  \caption{Boxplots showing the difference in empirical ($u=0.95$) tail dependence between ERA5 data and the probabilistic generative model. Note that $\lambda_{UL}$ and $\lambda_{LU}$ are grouped together in this case.}
  \label{fig:boxplot_taildep_nocv}
\end{figure}

\begin{figure}[H]
  \centering
  \includegraphics[scale=.45]{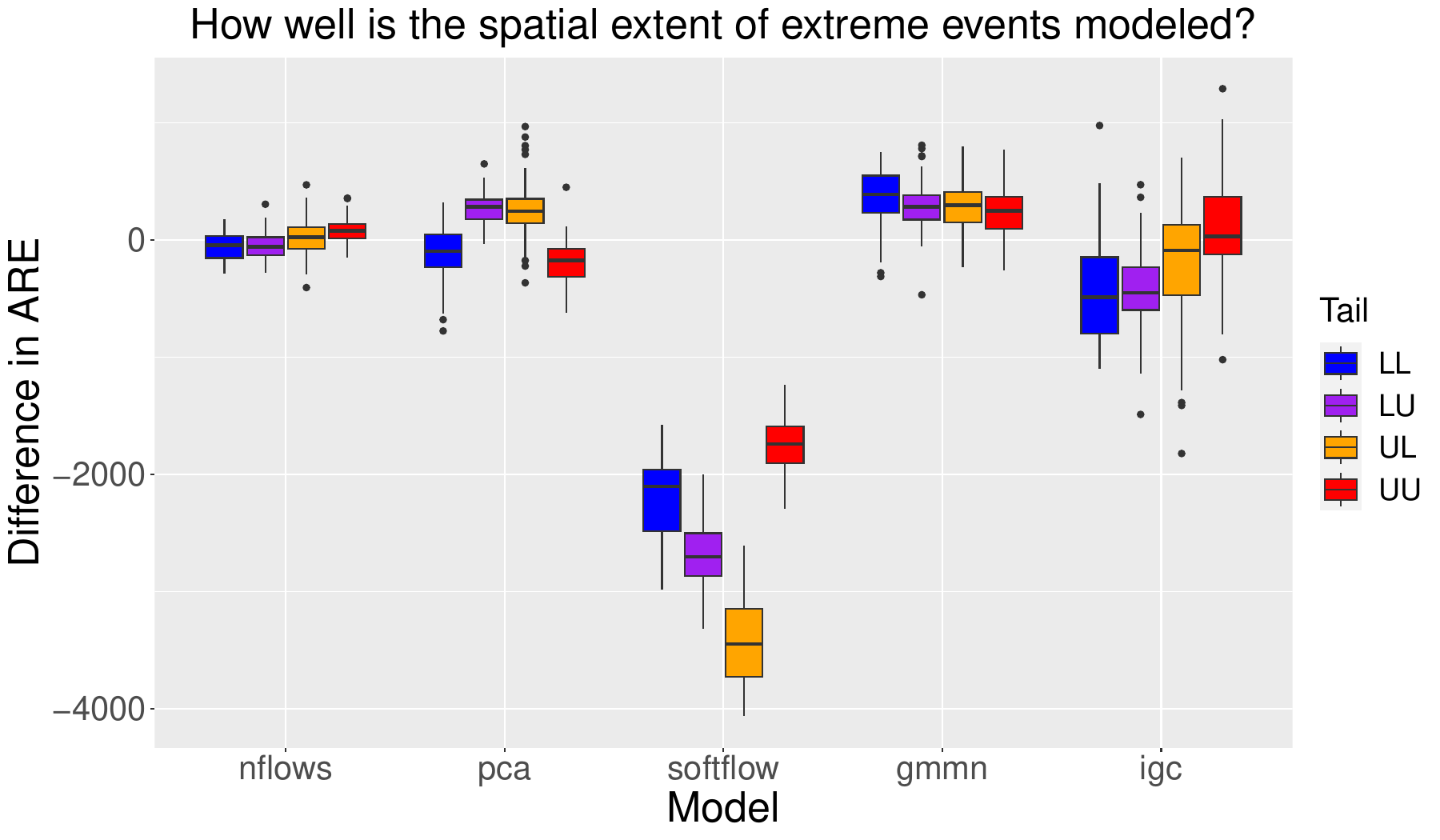}
  \caption{Boxplots showing the difference in empirical ($u=0.95$) ARE between ERA5 data and the probabilistic generative model.}
  \label{fig:boxplot_spatialextent_nocv}
\end{figure}

The following models were more competitive with the methods presented in the paper.
The Implicit Generative Copula \citep{janke2021} is a special type of Generative Moment Matching Network \citep{gmmn} that uses a novel neural network activation function to enforce marginal distributions to be approximately uniform during training.
SoftFlow \citep{Kim2020} was used to model tail dependence in COMET Flows \citep{mcdonald2022}. 
Our attempts to fit COMET Flows directly were not successful.
Figures \ref{fig:correlation_boxplot_nocv}-\ref{fig:boxplot_spatialextent_nocv} are analogous to Figures \ref{fig:cv_correlation_boxanddist}-\ref{fig:cv_spatialextent_boxanddist} but without any cross-validation; i.e.,\ the training and testing is performed on the entire ERA5 dataset.
Overall, models based on \texttt{nflows} are most accurate.
SoftFlow consistently overestimates the spatial extent of extremes.
Upon further inspection, this is due to poor marginal fitting---SoftFlow does not how to increase the probability extremes in the joint distribution, only in the marginal distributions. 
Finally, we found that the Generative Moment Matching Network was faster to train and more accurate than the Implicit Generative Copula, especially in terms of the estimated spatial correlation.

\begin{figure}[H]
  \centering
  \includegraphics[scale=.3]{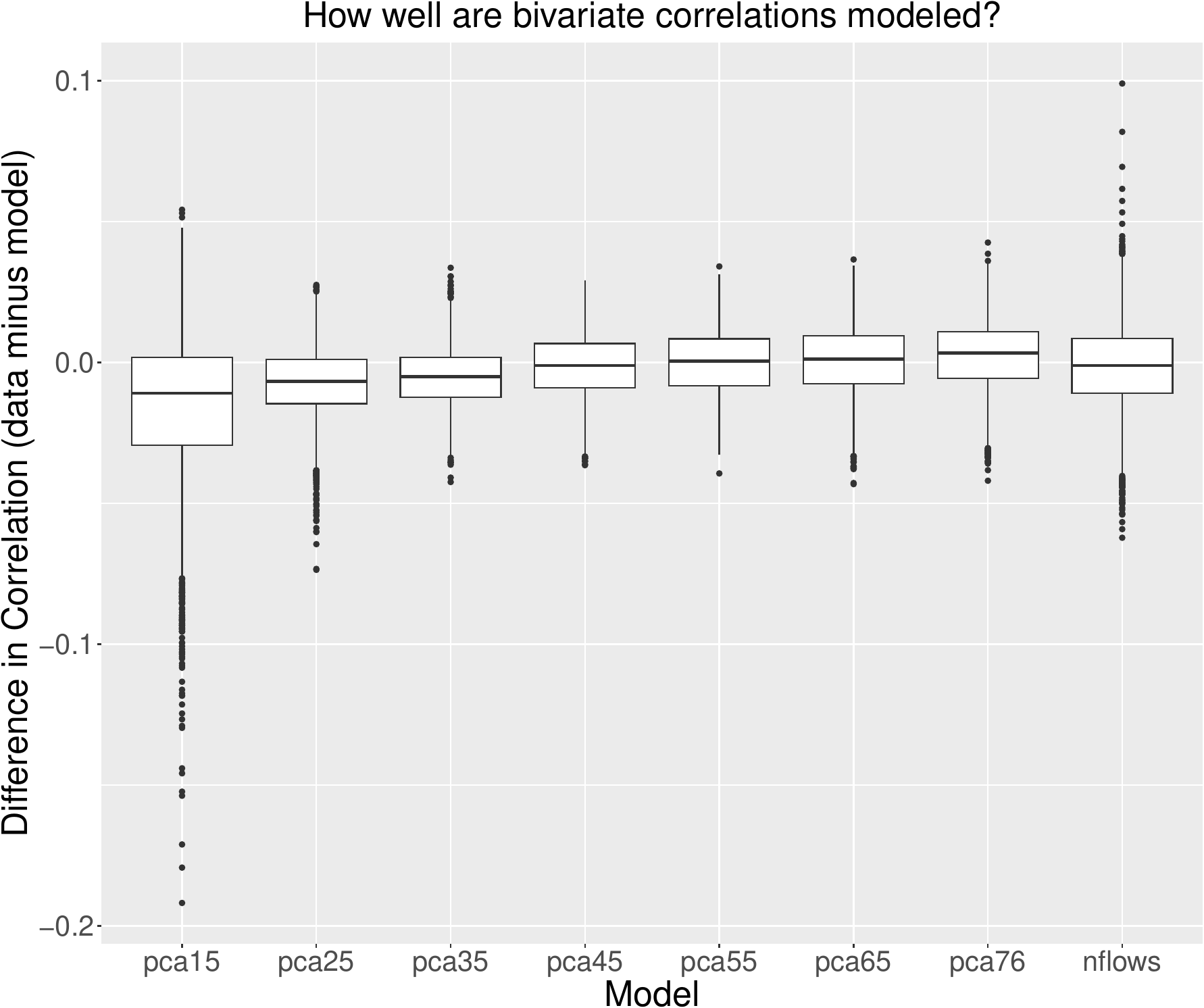}
  \caption{Increasing the number of principal components in Figure \ref{fig:cv_correlation_boxanddist}.}
  \label{fig:cv_correlation_boxanddist_pca}
\end{figure}

\begin{figure}[H]
  \centering
  \includegraphics[scale=.3]{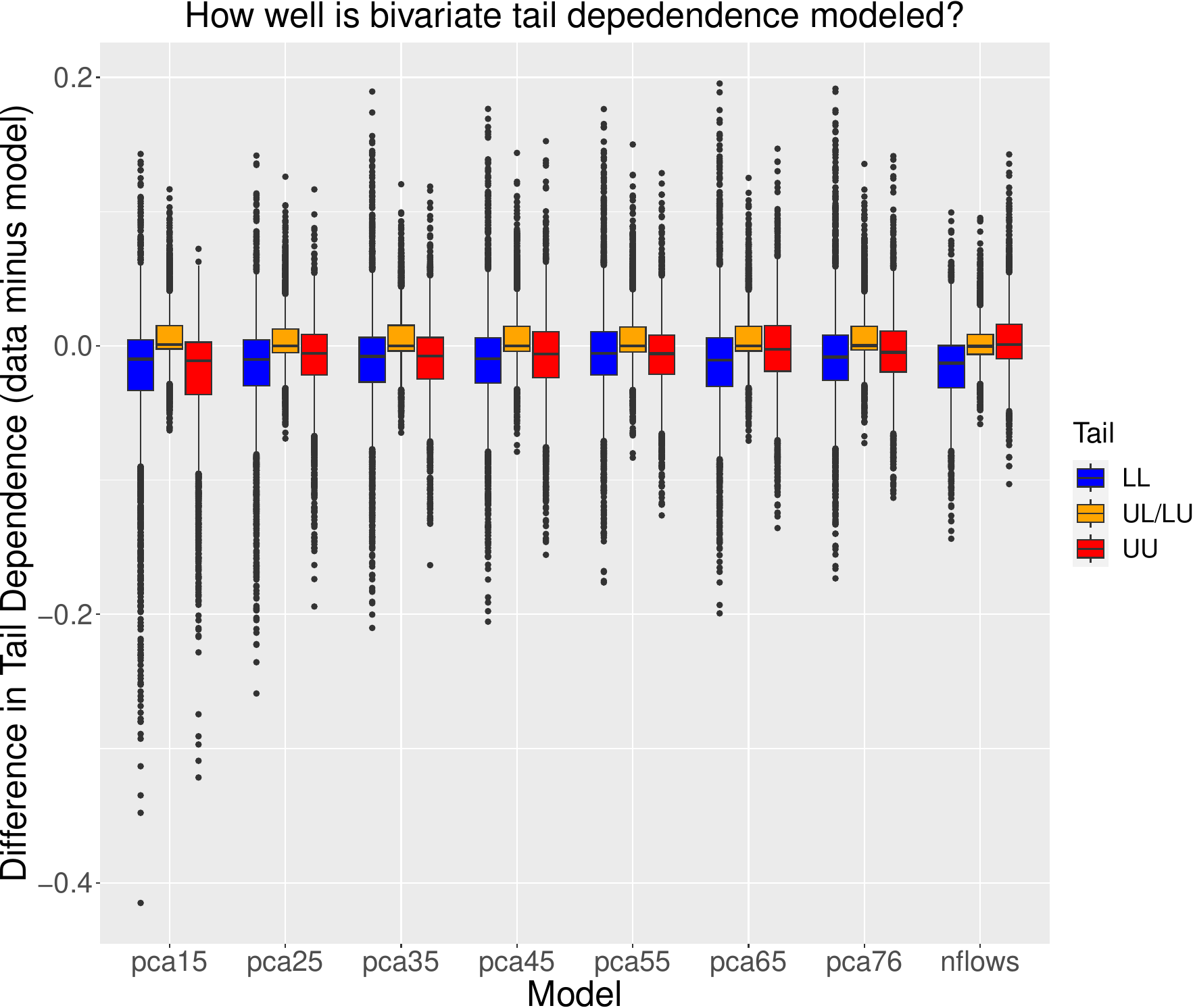}
  \caption{Increasing the number of principal components in Figure \ref{fig:cv_taildep_boxanddist}.}
  \label{fig:cv_taildep_boxanddist_pca}
\end{figure}

\begin{figure}[H]
  \centering
  \includegraphics[scale=.3]{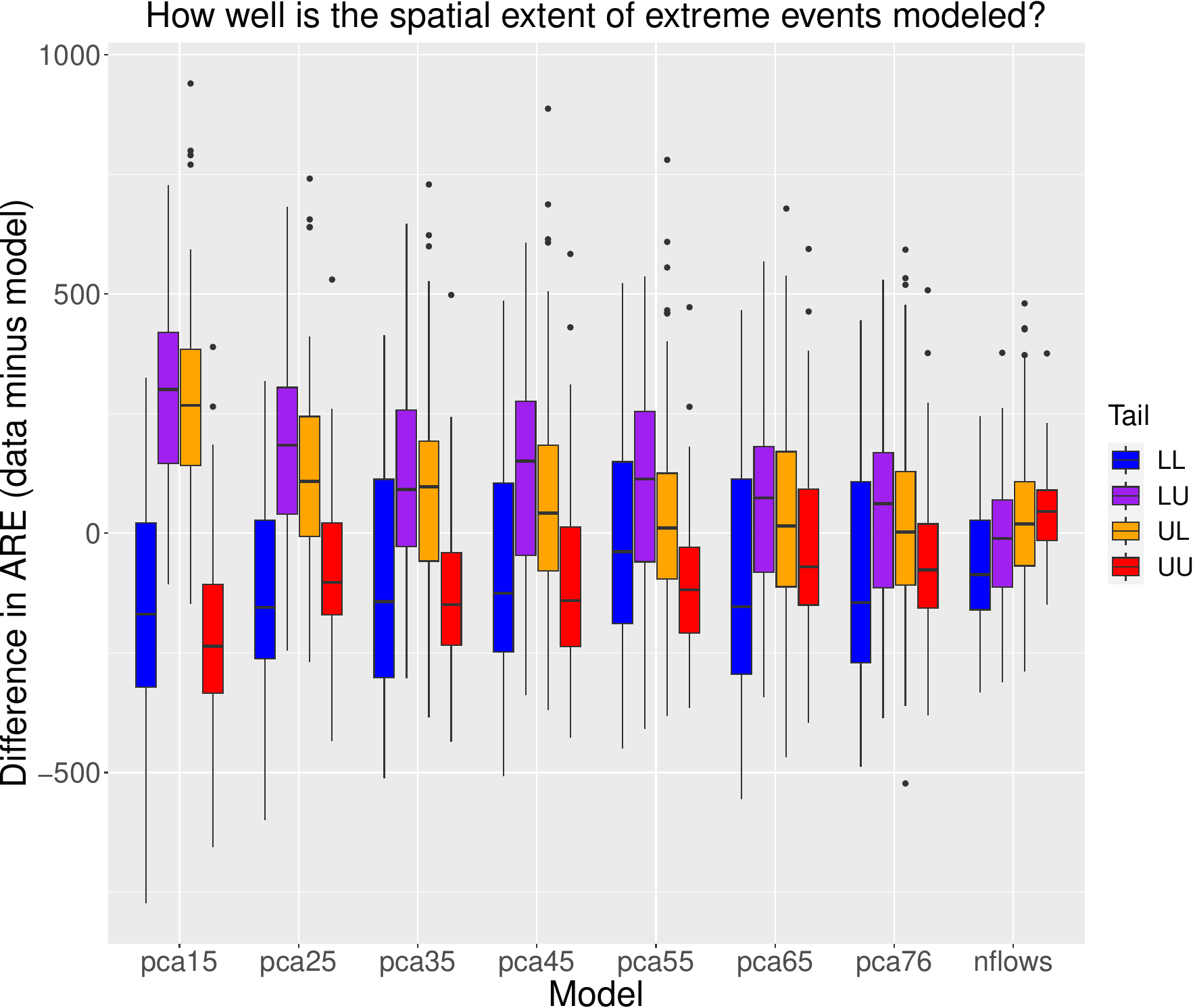}
  \caption{Increasing the number of principal components in Figure \ref{fig:cv_spatialextent_boxanddist}.}
  \label{fig:cv_spatialextent_boxanddist_pca}
\end{figure}

Figures \ref{fig:cv_correlation_boxanddist_pca}-\ref{fig:cv_spatialextent_boxanddist_pca} demonstrate how cross-validation performance improves as we increase the number of principal components in \eqref{eq:basisflow}.
Note that when all 76 principal components are used, the model is different from \texttt{nflows}, as normalizing flows are not invariant to linear transformations.

\section{Synthetic Data Analysis}

\label{app:synthetic}
Consider a bivariate student-$t$ distribution with $\nu=1$ degrees of freedom (i.e.,\ a bivariate Cauchy distribution) and dependence coefficient $\rho = 0.8$.
The tail dependence coefficients of this model are $\lambda_{UU} = \lambda_{LL} = f(\nu,\rho)$ and $\lambda_{UL} = \lambda_{LU} = f(\nu,-\rho)$, where $T_\nu(\cdot)$ denotes the student-$t$ cumulative distribution function with $\nu$ degrees of freedom and $f(\nu,\rho) = 2 T_{\nu+1} \left( - \sqrt{ \frac{ (\nu+1)(1-\rho)}{1+\rho} } \right)$.
Training data for our model consists of ${n_\text{dat}}=3940$ samples from this bivariate Cauchy distribution.
Although a sample size of 3940 would be considered small for most deep learning applications, it may be realistic for environmental data corresponding to climate records (e.g.,\ it is the sample size of the ERA5 reanalysis product used in Section \ref{sec:era5}).
We train the models for $20,000$ epochs using a batch size of 100 and learning rate of $0.0001$ in the ADAM minimizer; this also matches the settings in Section \ref{sec:era5}.

We briefly investigate the role of the input dimension of the noise vector $\mathbf{Z}$ to the generator in a \texttt{gmmn} model.
To match the framework of the normalizing flow, we consider the \texttt{gmmn} to take a two-dimensional noise vector $\mathbf{Z} \in \mathbb{R}^2$ as input.
We can also use a high-dimensional input space for $\mathbf{Z}$ by taking the same neural network architecture but removing the two-dimensional input and instead starting with the second, deeper layer.
Despite having less parameters, we see improved performance when $\mathbf{Z} \in \mathbb{R}^{100}$ rather than $\mathbf{Z} \in \mathbb{R}^2$.
As expected, tail dependence beyond the $1 - \frac{1}{3940}$ quantile is underestimated.
Without further parametric control, it will be difficult for probabilistic generative models to extrapolate beyond the in-sample tail behavior.

\begin{figure}[H]
  \centering
  \includegraphics[scale=.43]{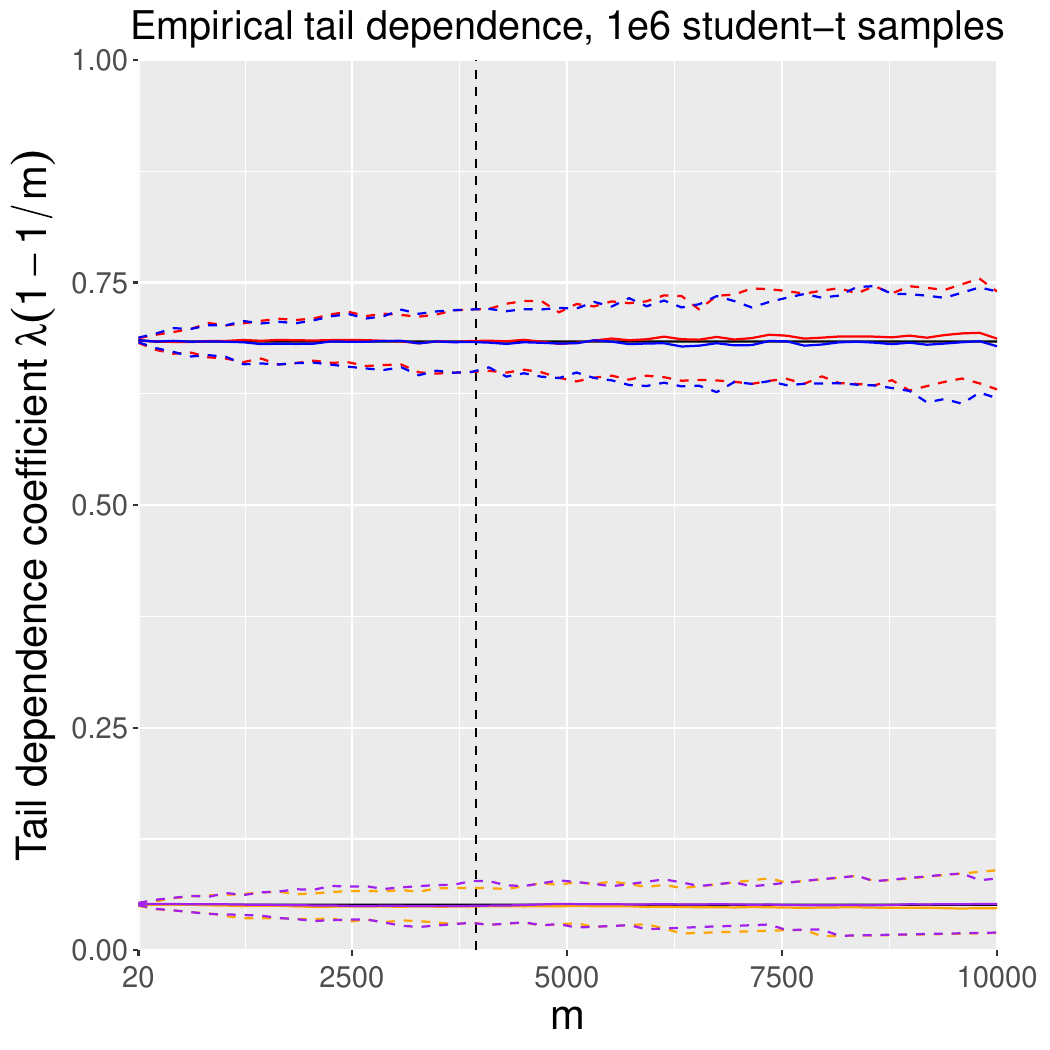}
  \includegraphics[scale=.43]{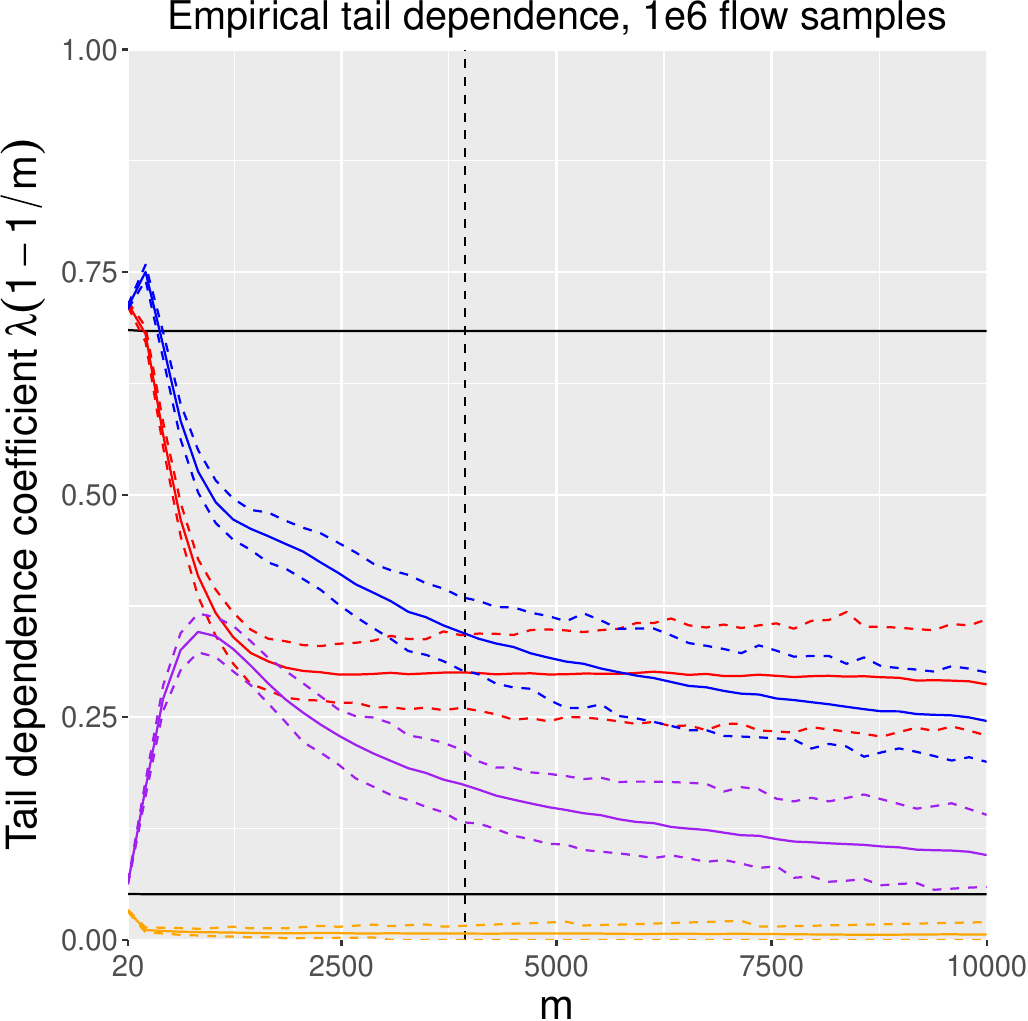}
  \includegraphics[scale=.43]{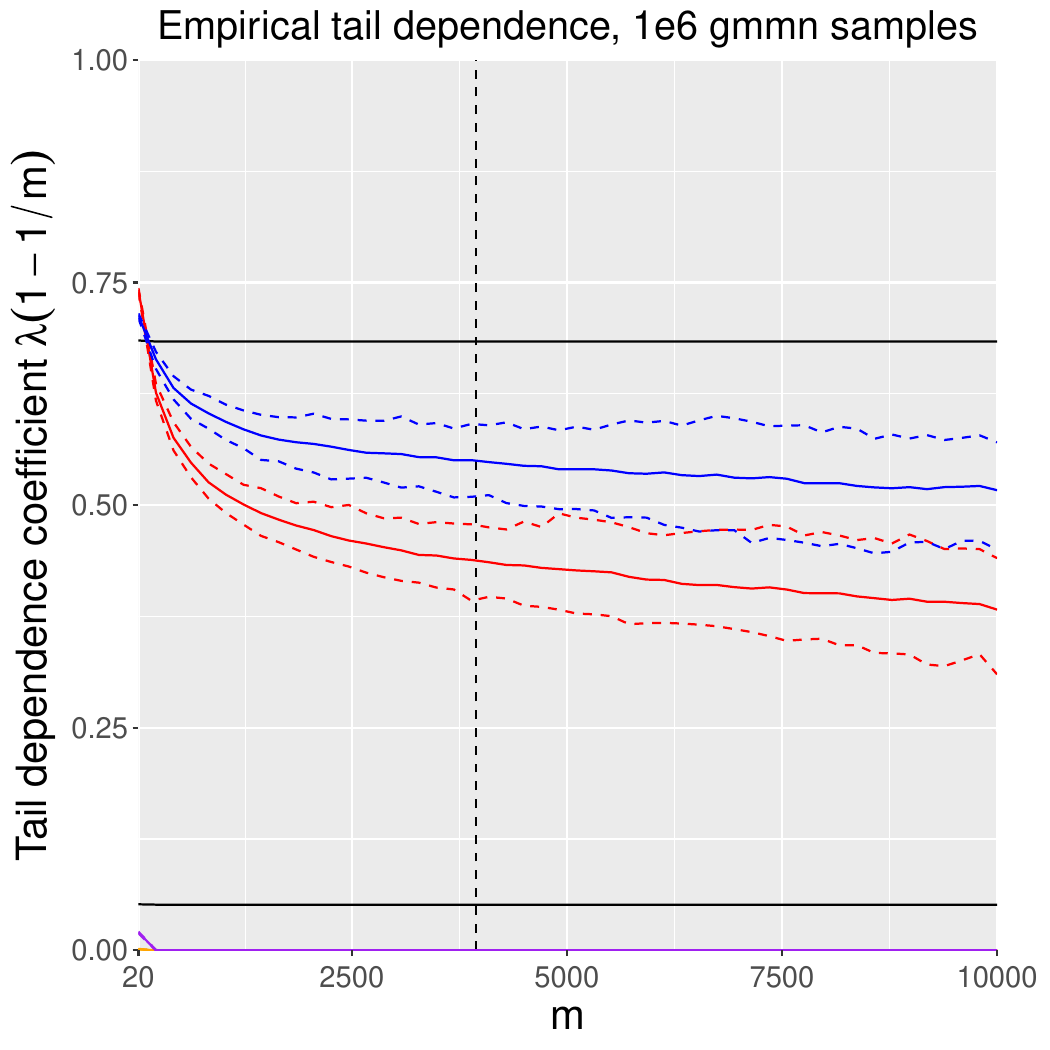}
  \includegraphics[scale=.43]{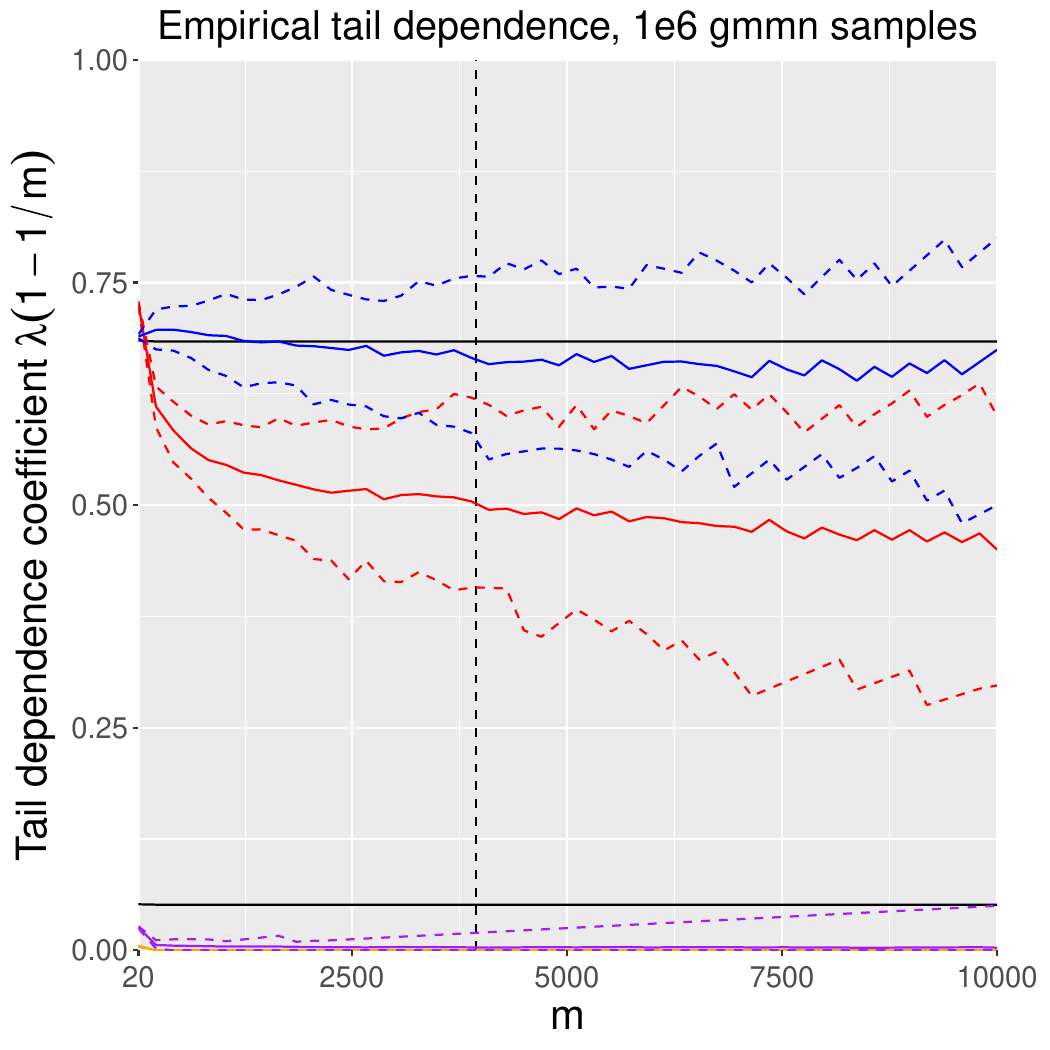}
  \caption{
    Displaying the estimated tail dependence coefficients and corresponding 95\% parametric-bootstrap confidence interval. 
    The top left panel shows the results under the true model.
    The top right panel shows the results for \texttt{nflows}, and the panels in the bottom row correspond to \texttt{gmmn} models.
    The two \texttt{gmmn} models have nearly the same number of parameters, but for the bottom left panel, the generator takes a 2-dimensional noise vector $\mathbf{Z}$ as input, while in the bottom right panel, $\mathbf{Z} \in \mathbb{R}^{100}$.
    Horizontal black lines show analytic tail dependence coefficients, and vertical black line shows the training sample size.
    Line colors: $\hat \lambda_{UU}$ red, $\hat \lambda_{LL}$ blue, $\hat \lambda_{UL}$ orange, $\hat \lambda_{LU}$ purple. Same axes on all plots.
  }
  \label{fig:studentttaildep}
\end{figure}

\clearpage

\newpage
\bibliographystyle{ametsocV6}
\bibliography{MLextremes}

\end{document}